\documentclass[onecolumn,traditabstract]{aa}
\usepackage{amssymb,amsmath,amsfonts}

\usepackage{txfonts}
\usepackage{bm}
\usepackage{graphicx, epsfig}

\usepackage{color}
\usepackage{subfigure}

\usepackage[english]{babel}

\newcommand{\be}{\begin{equation}}
\newcommand{\ee}{\end{equation}}
\newcommand{\bea}{\begin{eqnarray}}
\newcommand{\eea}{\end{eqnarray}}

\def\sf{S_{15}}

\newcommand\lsim{\mathrel{\rlap{\lower4pt\hbox{\hskip1pt$\sim$}}
    \raise1pt\hbox{$<$}}}
\newcommand\gsim{\mathrel{\rlap{\lower4pt\hbox{\hskip1pt$\sim$}}
    \raise1pt\hbox{$>$}}}

\def\dslash{\not{\hbox{\kern-2pt $\partial$}}}
\def\Dslash{\not{\hbox{\kern-4pt $D$}}}
\def\Oslash{\not{\hbox{\kern-4pt $O$}}}
\def\Qslash{\not{\hbox{\kern-4pt $Q$}}}
\def\pslash{\not{\hbox{\kern-2.3pt $p$}}}
\def\kslash{\not{\hbox{\kern-2.3pt $k$}}}
\def\qslash{\not{\hbox{\kern-2.3pt $q$}}}

\def\ee{\end{equation}}
\def\be{\begin{equation}}

\newcommand\bfd{{\bf d}}

\newcommand\bfv{{\bf v}}

\newcommand\bfN{{\bf N}}

\newcommand\bfC{{\bf C}}

\begin{document}

\title{Markov Chain Beam Randomization: a study of the
impact of PLANCK beam measurement errors on cosmological parameter estimation
}
\titlerunning{\texttt{MCBR}: a study of the impact of PLANCK beam measurement errors on cosmological parameter estimation}
\author{
G.~Rocha\inst{1,2}
\and
L.~Pagano\inst{3}
\and 
K.M.~G\'{o}rski\inst{1,2,4} 
\and
K.M.~Huffenberger\inst{5} 
\and
C.R.~Lawrence\inst{1} 
\and
A.E.~Lange\inst{2}
}
\institute{
Jet Propulsion Laboratory, California Institute of Technology, 4800 Oak Grove Drive, Pasadena CA 91109, U.S.A.
\and
California Institute of Technology, Pasadena CA 91125, U.S.A.
\and
Physics Department and sezione INFN, University of Rome ``La Sapienza'', Ple Aldo Moro 2, 00185 Rome, Italy.
\and
Warsaw University Observatory, Aleje Ujazdowskie 4, 00478 Warszawa, Poland.
\and
Department of Physics, University of Miami, 1320 Campo Sano Avenue, Coral Gables, FL 33124.
}

\date{Received date / Accepted date}
\abstract{We introduce a new method to propagate uncertainties in the beam shapes used to measure the cosmic microwave background to cosmological parameters determined from those measurements.  The method, called Markov Chain Beam Randomization (\texttt{MCBR}), randomly samples from a set of templates or functions that describe the beam uncertainties.  The method is much faster than direct numerical integration over systematic `nuisance' parameters, and is not restricted to simple, idealized cases as is analytic marginalization.  It does not assume the data are normally distributed, and does not require Gaussian priors on the specific systematic uncertainties.  We show that  \texttt{MCBR} properly accounts for and provides the marginalized errors of the parameters.  The method can be generalized and used to propagate any systematic uncertainties for which a set of templates is available.   We apply the method to the Planck satellite, and consider future experiments.  Beam measurement errors should have a small effect on cosmological parameters as long as the beam fitting is performed after removal of $1/f$ noise.
}


\maketitle

\section{Introduction} \label{introduction}

Observations of the cosmic microwave background (CMB) can be interpreted only in light of a detailed knowledge of the angular response of the instrument to radiation, i.e., the shapes of the `beams.' It is almost always the case that the beams from single-aperture telescopes (but not interferometers) can be approximated as two-dimensional Gaussians. It is never the case that a gaussian approximation provides an adequate description of the beams of an experiment that measures the CMB with high signal-to-noise ratio. If the beams were known perfectly, their effects on the data could be calculated perfectly, if painfully. Unfortunately, beams are never known perfectly, and among the outstanding issues for any CMB experiment are how to optimize the beams in the first place, and how to control and account for beam uncertainties in the data analysis.

The effects of beam uncertainties can be analyzed in maps, power spectra, and cosmological parameters determined from the data. Each has benefits. Because cosmological parameters are a key product of any experiment, and because they are sensitive to extremely small effects impossible to detect pixel by pixel, they are particularly valuable. Historically, however, calculation of the effects of beam uncertainties on cosmological parameters has been done either analytically, which requires over-simplified beam shapes, or numerically, at great computational cost.

We introduce in this paper a method for calculating the effects of beam uncertainties on cosmological parameters determined from CMB observations that is both fast and flexible. It requires only that beam uncertainties, or for that matter any other systematic effect, can be represented by a set of functions or templates, which could be obtained from Monte Carlo simulations. It does not assume that the data themselves are Gaussian-distributed, or that the uncertainties have Gaussian priors.

In \S\,II we describe the method, called Markov Chain Beam Randomization or \texttt{MCBR}, and we show that the \texttt{MCBR} technique produces correct marginalized errors. \S\,III summarizes the beam fitting procedure developed in a previous paper (\cite{kevinetal09}). \S\,IV describes the implementation of \hbox{MCBR}.
 In \S\,VI we apply the method to the Planck experiment, and consider future experiments.

\section{\texttt{MCBR}: Markov Chain Beam Randomization \label{mcbr}}

In the past, marginalization over systematic parameters has been carried out either numerically or analytically (\cite{bridle02}); both methods are currently implemented in  \texttt{cosmomc} (\cite{Lewis:2002ah}).
Assuming likelihoods are Gaussian one typically has a marginalization of the form:

\be
L \propto \int d \alpha P(\alpha) \exp [ -(\alpha \bfv - \bfd)^{T} \bfN^{-1} (\alpha \bfv - \bfd) /2 ]
\ee
where $\bfd$ is the theoretical (predicted) data 
minus the observed data and  $\alpha \bfv$ is an approximate template describing the correction applied to the predicted data due to systematics,  $\bfN$ is the noise covariance matrix, and $P(\alpha)$ is the prior.
The marginalization is given by:
\be
-2 \ln L=  \bfd^{T} \left( \bfN^{-1} - \frac{\bfN^{-1} \bfv \bfv^{T} \bfN^{-1}}{\bfv^{T} \bfN^{-1} \bfv}  \right) \bfd  + \ln ( \bfv^{T} \bfN^{-1} \bfv ) + \text{c}
\ee
where $\text{c}$ is a constant. If $\bfv$ is independent of the data and parameters then $L \propto e^{-\chi_{eff}^{2}/2}$, with:
\be
\chi_{eff}^{2} = \bfd^{T} \left( \bfN^{-1} - \frac{\bfN^{-1} \bfv \bfv^{T} \bfN^{-1}}{\bfv^{T} \bfN^{-1} \bfv} \right) \bfd = \chi_{best-fit}^{2}
\ee

In the case of beam uncertainties, the analytic approach is feasible only if the beams are assumed to be Gaussian.  This is not realistic.

It is customary to characterize anisotropies in the Cosmic Microwave Background by their angular  power spectrum, $\bfC_{\ell}$ for both temperature and polarization. $\bf{C}_{\ell}$ is a $3 \times 3$ matrix for T (temperature) and E or B (grad-type or curl-type polarization):
\begin{equation}
{\mathbf{C}}_{\ell} =
\left(
\begin{array}{ccc}
{C}_{\ell}^{TT} & {C}_{\ell}^{TE} & {C}_{\ell}^{TB} \\
{C}_{\ell}^{TE} & {C}_{\ell}^{EE} & {C}_{\ell}^{EB}  \\
{C}_{\ell}^{TB} &   {C}_{\ell}^{EB} & {C}_{\ell}^{BB} \\
\end{array}
\right),
\end{equation}
Hereafter, for the sake of simplicity, most equations will refer to the angular power spectrum, $C_{\ell}$, for a single component, say temperature.
The telescope beam smooths the anisotropies, supressing power at higher multipoles.  We refer to the ratio of the measured power spectrum of the sky and our true power spectrum as the transfer function, ${\cal{B}}_{\ell}=B_{\ell}^{2}$. Here we assume the beam transfer functions are the same for temperature and polarization.

To obtain unbiased estimates of the parameters that characterize the cosmology, we must repair this suppression based on knowledge of the beam.  Uncertainties in the beam propagate into uncertainties in the cosmological parameters.

 We assume that the beam uncertainties can be described by a set of functions or templates, taken here to be the set of transfer functions obtained by the beam fitting procedure described in \S\,~\ref{beam-fitting}. These templates are given in multipole space  by:
\be
{\cal{B}}_{\ell}^{r} =  (B_{\ell}^{r})^{2} =  (B_{\ell} \times r_{\ell})^{2}
\ee
where the ratios $r_{\ell}$ represent the possible deviations from the true fiducial beam.
We choose the beam transfer function randomly from the set of $N$ simulations (here $N=1280$)  for each step of the Markov Chain Monte Carlo when probing the cosmological parameters space. This means that at each step of the chain the theoretical power spectrum, $\bfC_{\ell}$, is multiplied by the randomly chosen beam, ${\cal{B}}_{\ell}^{r}$.  
We assume all transfer functions in the set are equally probable. 
 
To estimate constraints on cosmological parameters, we need to compare the model with the data via a chosen Likelihood and an algorithm to sample cosmological parameters.  Here we make use of the package \texttt{cosmomc}. To incorporate \texttt{MCBR} we modify \texttt{cosmomc} to enable the usage of  a random ${\cal{B}}_{\ell}^{r}$ for each theoretical model generated with \texttt{CAMB} (\cite{Lewis:1999bs}) or \texttt{PICO} (\cite{Fendt:2006uh}).  
This is done by modifying the \texttt{cmbdata} module of the \texttt{cosmomc} code. 

We start by creating simulated datasets with noise properties specific to the instrument under consideration, in our case Planck and an example of a future experiment (see \S\,~\ref{beam-analysis}). These simulated datasets are given in terms of the angular Power Spectrum $C^{obs}_{\ell}$:

\be
C^{obs}_{\ell} = C^{\text{wmap}}_{\ell} {\cal{B}}_{\ell} + \cal{N}_{\ell} \label{Clobs}
\ee
where $\cal{N}_{\ell}$ is the noise power spectrum and ${\cal{B}}_{\ell}= B_{\ell}^{2}$ is the beam transfer function and $C^{\text{wmap}}_{\ell}$ is the $\Lambda$ CDM spectrum best-fitting current WMAP data.
In the case of a symmetric Gaussian beam, 
$B(\bf{x}) = \frac{1}{2\pi \sigma^2} \exp\left\{\frac{-|\bf{x}|^2}{2\sigma^2}\right\}$, so that $B_{\ell} =e^{-\frac{1}{2}\sigma^2\ell^2}$.  However, the Planck beams are not adequately represented by Gaussians.  Instead, we use realistic beams calculated from a full diffraction analysis of the telescope using GRASP9  (\cite{sandri:2002,sandri:2009, mafei:2009,yurchenko:2004}).

As our purpose here is to introduce and validate the \texttt{MCBR} method it suffices to assume full-sky coverage. 
Considerations of realistic complications (such as cut-sky, foregrounds, etc.) is deferred to a future publication. Our purpose here is to establish the relative importance of propagating beam errors to cosmological parameters rather than to make comprehensive predictions for Planck.
Hereafter to compare the observed dataset,
 $\bfC^{obs}_{\ell}$, with theoretical models we use the exact full-sky likelihood (with $\hat{\bfC}_\ell = {\bfC}^{obs}_{\ell}$) (\cite{DA/BJK00}):

\begin{equation}
\label{exactfullsky}
- 2\ln L(\hat{\bf{C}_\ell} | \bf{C}_\ell)
 = (2\ell + 1) \left( 
\ln |\bf{C_\ell}| + 
\mathrm{Tr} \left(
 \hat{\bf{C}_\ell} \bf{C}_\ell^{-1} \right)
\right),
\end{equation}
 i.e., the Inverse Wishart distribution for Temperature and Polarization. 
In \texttt{cosmomc} this distribution is coded in function \texttt{ChiSqExact} (\cite{Lewis:2005tp}).
We analyse these datasets with a modified version of this function, built to include the \texttt{MCBR} procedure in the code.

The $C_{\ell}$ of the theoretical model is given by:

\begin{equation} 
\tilde{C}_{\ell}= C_{\ell} \times {\cal{B}}_{\ell}^{r} 
\label{clcor}
\end{equation}
where ${\cal{B}}_{\ell}^{r}$ is the randomly chosen transfer function.
To incorporate both the beam and the uniform white noise in the likelihood expression one should replace:

\begin{eqnarray}
\centering
 \hat{C}_{\ell} &\rightarrow&  C^{obs}_{\ell} \label{Clobs2} \\
C_{\ell} &\rightarrow& C_{\ell}^{th} \times {\cal{B}}_{\ell}^{r} + \cal{N}_{\ell} \label{Clth-bn}
\end{eqnarray}
where $C_{\ell}^{obs}$  is given by Equation~\ref{Clobs}, $C_{\ell}^{th}$ is the theoretical power
spectrum computed e.g. by CAMB, and $B_{\ell}^r$ is the randomly chosen beam transfer function.

In the \texttt{MCBR} scheme, sampling of the beam templates is equivalent to sampling from the proposal distribution. The Metropolis-Hastings algorithm accepts the move from $\theta_{n}$ to $\theta_{n+1}$  in the Markov chain by evaluating the ratio:
\be
\frac{P(\theta_{n+1}) q(\theta_{n+1},\theta_{n})}{P(\theta_{n}) q(\theta_{n},\theta_{n+1})}
\label{mh}
\ee
where $P$ is the posterior distribution we wish to sample from and $q$ is the proposal distribution.
We draw the proposal at position $\theta_{n}$  of the parameter space from $q(\theta_{n+1},\theta_{n})$.
Here  $\theta = (\theta_{cp}, \theta_{b}) $, with $\theta_{cp}$ the subset of cosmological parameters and $\theta_{b}$ the beam parameter.
The joint  proposal density for $\theta $ factors into
\be
q(\theta_{n+1},\theta_{n}) = q_{cp} (\theta_{cp, n+1}, \theta_{cp, n}) q_b(\theta_{b, n+1},\theta_{b,n}),
\ee

where $\theta_{cp, n+1}$ refers to $(\theta_{cp})_{n+1}$ and  $\theta_{b, n+1}$  to $(\theta_{b})_{n+1}$.
Now, we take the $q_b(\theta_{b, n+1}, \theta_{b,n})$ to be the posterior distribution of the beam parameters given the beam fitting data (in our case the Jupiter beam fitting data (see \S\,~\ref{beam-fitting})), i.e., 
\be
q_b( \theta_{b, n+1},\theta_{b,n}) = P_b( \theta_{b, n} | \text{beamdata} ) 
\ee
Furthermore
\be
P(\theta_{n+1}) = P_{cp}( \theta_{n+1} | \text{mapdata}) P_{b}( \theta_{b, n+1} | \text{beamdata}) 
\ee
(for instance in our study here $P_{cp}( \theta | \text{mapdata}) = L(\text{mapdata}| \theta_{cp}, \theta_{b}) p_{cp}(\theta_{cp})$  where $L$ is the Likelihood given in Equation~\ref{exactfullsky} and $p_{cp}$ the prior on cosmological parameters.)
Hence the ratio in Equation~\ref{mh} becomes:
\be 
\frac{ P_{cp}( \theta_{n+1} ) q_{cp}( \theta_{cp, n+1}  , \theta_{cp, n} ) } { P_{cp}( \theta_{n} ) q_{cp}( \theta_{cp, n} ,  \theta_{cp, n+1} ) } 
\ee
as  $P_{b}$ and $q_{b}$ cancel out.

Hence random sampling from the set of beam templates at each step of the Markov chain is equivalent to sampling from a proposal density  that, by construction, is identical to the posterior distribution of the beam parameters given the beam fitting data.

To illustrate how the \texttt{MCBR}  procedure works, we give here the steps followed in our analysis (see section~\ref{results}).
We start by comparing two cases:

\begin{enumerate}
\item \texttt{cosmomc} run with the `true' fiducial beam transfer alone, ${\cal{B}}_{\ell}$.
\item \texttt{cosmomc} run with the \texttt{MCBR} procedure for the set of  beam transfer functions, ${\cal{B}}_{\ell}^{r}$, obtained from the beam fitting step. 
\end{enumerate}

To this end:
\begin{itemize}
\item We generate a simulated data set using as fiducial the `true' beam transfer ${\cal{B}}_{\ell}$
\item We analyse this simulated data set with \texttt{cosmomc}, including in the code just the effect of the ${\cal{B}}_{\ell}$ (ie the theoretical $C_{\ell}$ is multiplied by the true beam transfer, ${\cal{B}}_{\ell}$)
\item  We analyse this simulated data set with a modified version of \texttt{cosmomc} in which the theoretical $C_{\ell}$ is multiplied by the randomly chosen beam,
 ${\cal{B}}_{\ell}^{r}$ at each step of the chain---i.e., with in-built \texttt{MCBR}
\end{itemize}

The MCBR method can be used to propagate any systematic uncertainties that can be characterized by a set of templates.  We turn these into multiplicative and additive corrections to the ${C}_{\ell}$, encode the corrected $C_{\ell}$ into the likelihood, and randomly sample from the set of templates at the Markov Chain Monte Carlo step of parameter estimation.   The data do not have to be normally distributed.  Furthermore, unlike analytic marginalization, the method does not require Gaussian priors on the uncertainties.

\subsection{Validation}\label{validation}

To demonstrate that that the \texttt{MCBR} technique gives correct marginalized errors, we compared the results given by  \texttt{MCBR}  to those from a `brute force' \texttt{cosmomc} calculation in which the beam was taken as another parameter.  
We did this for three simulated datasets, the first generated using the `true' beam transfer function ${\cal{B}}_{\ell}$, the second and third using beam transfer functions that were chosen to be mildly and extremely far from the true one, respectively.   We simplified the test cases by assuming that the beam was a symmetric $7'$ (FWHM) Gaussian, with 4\% variations of the fwhm of the beam.

The brute force calculation was done by probing the beam parameter space in \texttt{cosmomc} in the same way as for any of the other parameters, and considering the default proposal density already implemented in \texttt{cosmomc}.
The beam parameter is included by transforming the theoretical $C_{\ell} (\theta_{cp})$ output by CAMB at each Markov chain step into
\be
C_{\ell}(\theta_{cp}, \theta_{b}) = C_{\ell}(\theta_{cp}) \times B_{\ell}^{n} + {\cal{N}}_{\ell},
\ee
where $\text{fwhm}_{n}$ is the width of $B_{\ell}^{n}$, the Gaussian beam currently sampled.  This theoretical $C_{\ell}(\theta_{cp}, \theta_{b})$ is used in the Likelihood expression. To move the Markov chain to the next position in parameter space we use the default proposal density in \texttt{cosmomc}, usually an $N$-d Gaussian. 
The proposed new point is $\text{accepted} / \text{rejected}$ following the same prescription used for the other cosmological parameters.  A final marginalized distribution of the beam is output along with the other cosmological parameter constraints.

For the \texttt{MCBR} calculation we analysed the simulated data with a modified version of \texttt{cosmomc} in which the theoretical $C_{\ell}$ is multiplied by the randomly chosen beam transfer,
 ${\cal{B}}_{\ell}^{r}$ at each step of the chain.

\begin{figure*}
\begin{center}
\includegraphics[scale=0.75]{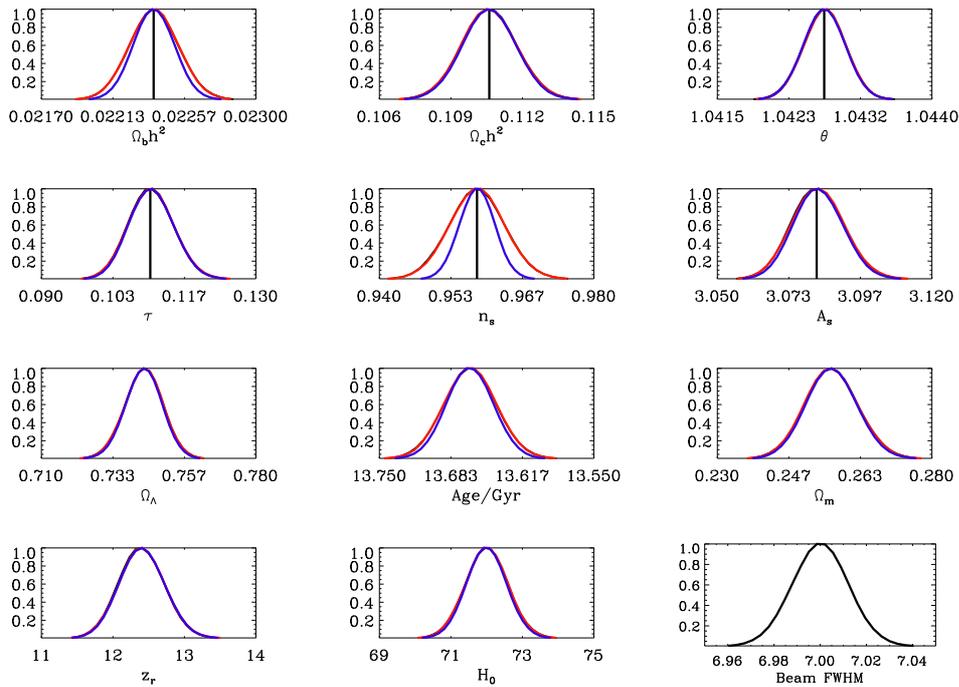}
\caption{Marginalized parameter constraints for Planck 143 GHz with $7^{'}$ beam with $4\% $ variations,
 for the analysis with the `true' reference fiducial beam using beamparameter approach (black) 
and \texttt{MCBR} (red), the blue line is the analysis o f the same dataset without including the beam uncertainty.}
\label{fig:gauss_comp}
\end{center}
\end{figure*}

\begin{figure*}
\begin{center}
\includegraphics[scale=0.75]{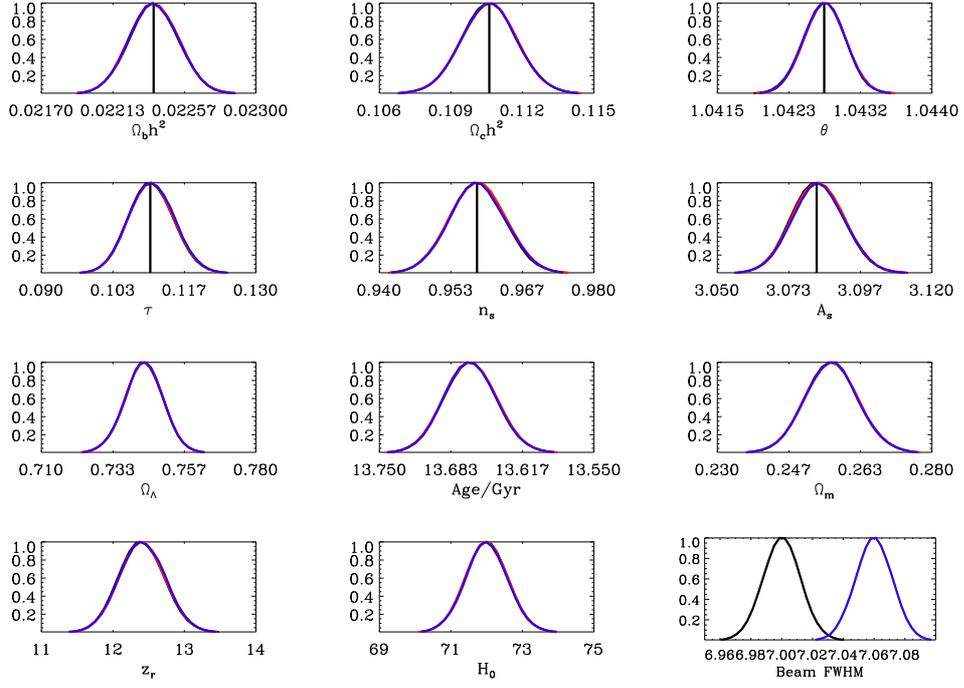}
\caption{Marginalized parameter constraints for Planck 143 GHz with $7^{'}$ beam with $4\% $ variations, 
for the analysis with the `true' reference fiducial beam (black) and for the mildly deviated beam transfer, ${\cal{B}}_{\ell}^{mild} = (B_{\ell} \times r^{mild}_{\ell})^{2}$, using beamparameter approach (blue)
 and \texttt{MCBR} (red), both beamparameter and \texttt{MCBR} give same distributions}
\label{fig:gauss_compfunc}
\end{center}
\end{figure*}

\begin{figure*}
\begin{center}
\includegraphics[scale=0.75]{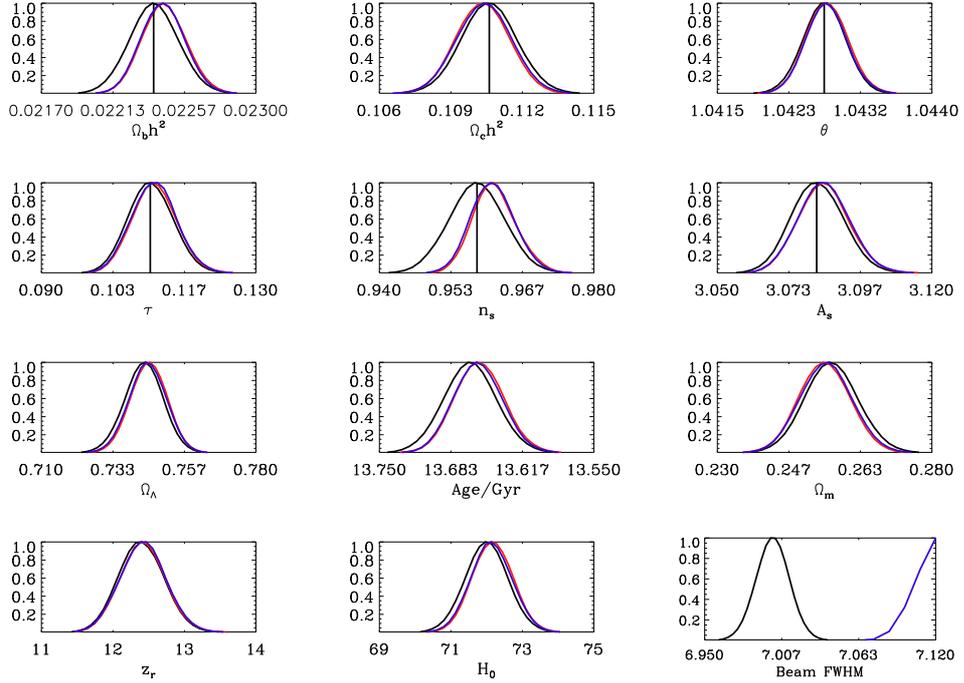}
\caption{Marginalized parameter constraints for Planck 143 GHz with $7^{'}$ beam with $4\% $ variations,  for the analysis with the `true' 
reference fiducial beam (black) and for the extremely deviated fiducial beam,
 ${\cal{B}}_{\ell}^{ext} = (B_{\ell} \times r^{ext}_{\ell})^{2}$, using beamparameter approach (blue) and \texttt{MCBR} (red), both beamparameter and \texttt{MCBR} give same distributions.}
\label{fig:gauss_compfunc_ext}
\end{center}
\end{figure*}

The results are plotted in Figures~\ref{fig:gauss_comp}, \ref{fig:gauss_compfunc}, and \ref{fig:gauss_compfunc_ext}.
In all cases, we find same parameter distributions for both methods.  As expected, the extreme deviated beam results in a biased estimation of parameters, especially $n_{s}$, but equally for both the `beamparameter' and the \texttt{MCBR} procedures.
 
\section{Beam fits and transfer function ensembles \label{beam-fitting}}

\begin{figure*}
  \includegraphics[width=0.49\linewidth]{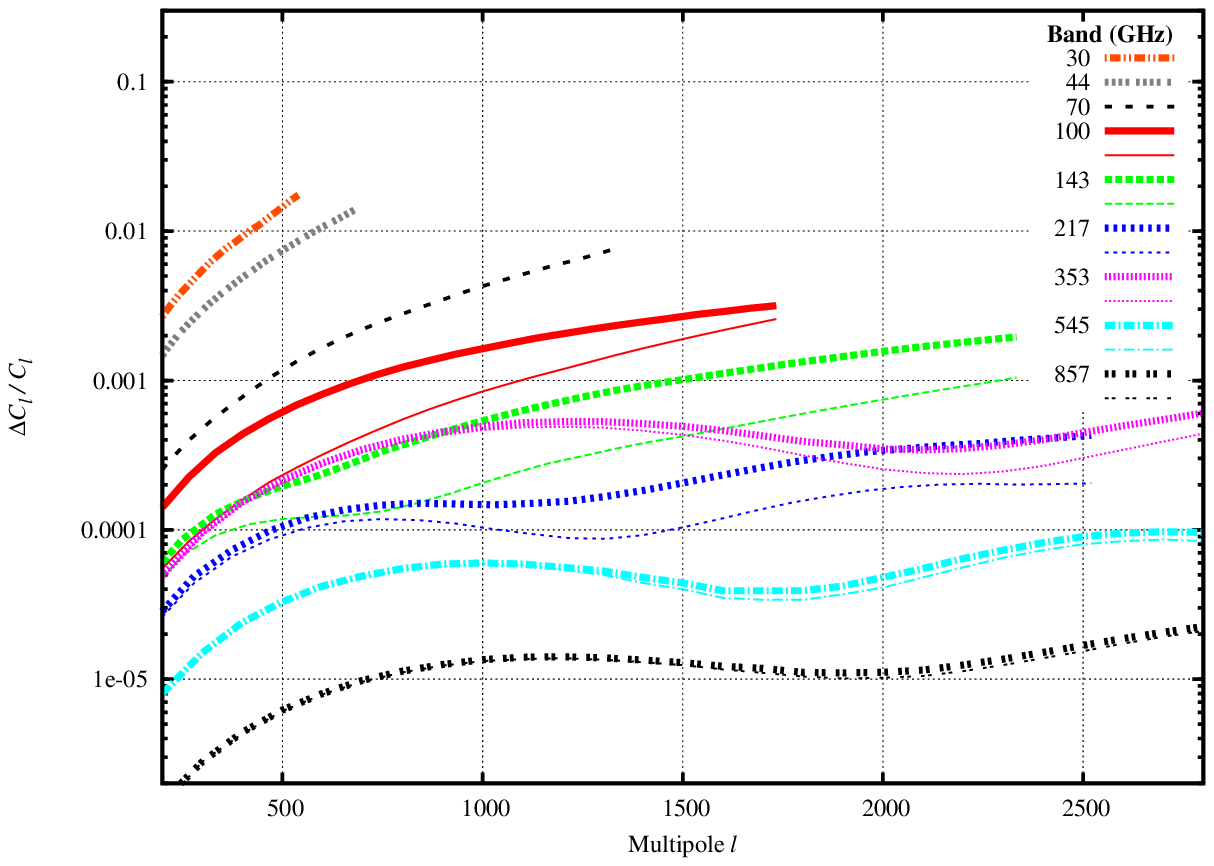}
  \includegraphics[width=0.49\linewidth]{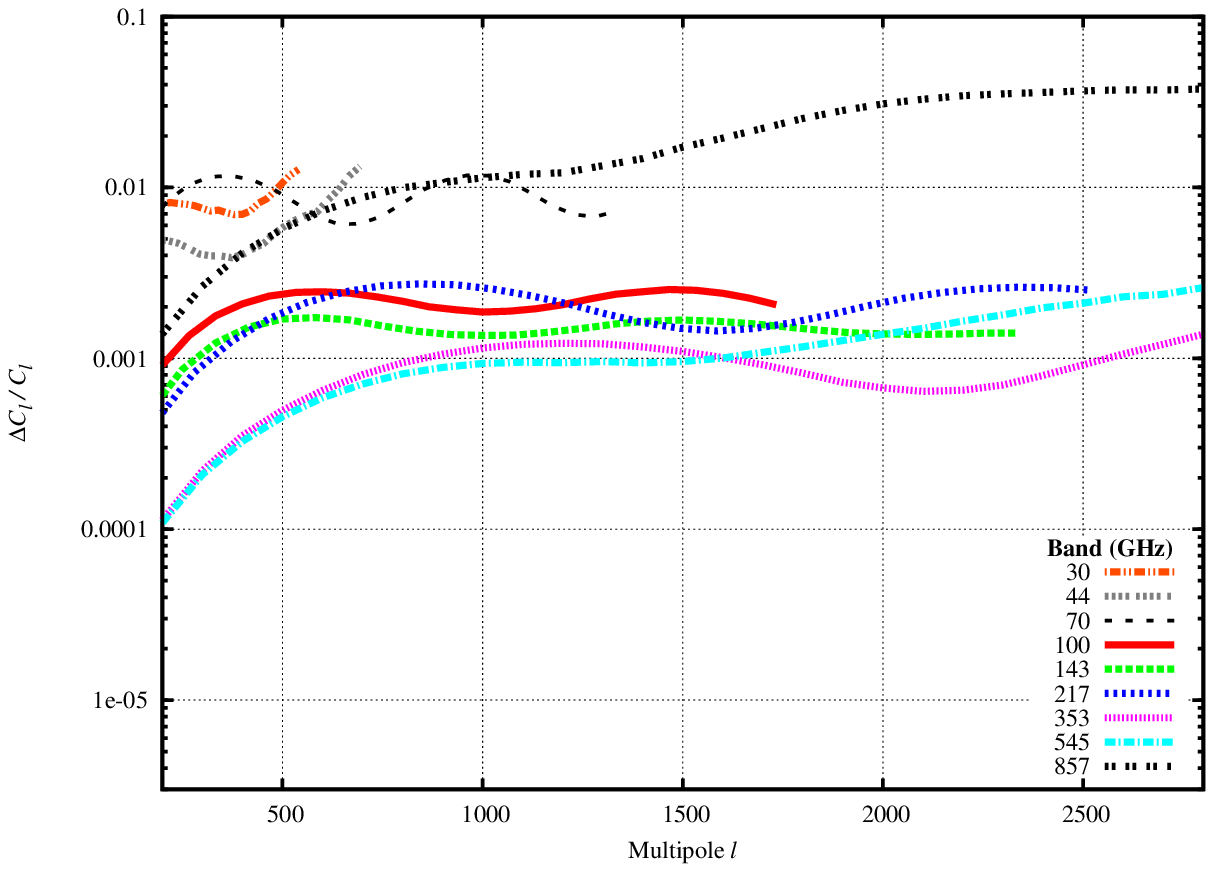}
\caption{At each multipole, $68\%$ of the fitted Monte Carlo transfer functions recover spectra closer to the true power spectrum than the indicated line.  Left: parametric model. Right: non-parametric model based on orthogonal functions, where the flexibility requires less knowledge of the beam, but yields larger errors.}
\label{fig:ratiosk}
\end{figure*}

\begin{figure}[t]
\begin{center}
  \includegraphics[width=0.45\linewidth]{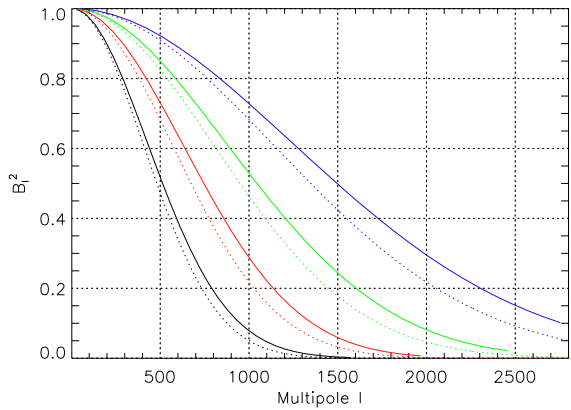}
\caption{Nominal Gaussian blue-book beams (dotted line) vs Fiducial realistic Grasp9 beams based on a Gaussian fit  (solid line) for 70GHz (black), 100GHz (red), 143GHz (green) and 217GHz(blue).}
\label{fig:bluegrasp}
\end{center}
\end{figure}

We characterize the beam uncertainty for Planck with a Monte Carlo ensemble of transfer functions (\cite{kevinetal09}) generated by repeated simulation of Jupiter observations using the detector noise and pointing errors expected before flight.  Each realization yields a representative transfer function.  The beams are calculated with GRASP9 (\cite{sandri:2002,sandri:2009,mafei:2009,yurchenko:2004}), and we employ two methods of beam reconstruction to reproduce them from the planet scans.  The first uses a rigid linearized parametric model; the second expands the beam in orthogonal functions (see \cite{Rocha01} for a previous application of such functions in CMB analysis). 
Figure~\ref{fig:bluegrasp} shows the nominal Gaussian beams with blue-book fwhm values to that of the fiducial realistic Grasp beams based on a Gaussian fit (see table\ref{tab:exp}).  
From the beam reconstruction procedure presented in \cite{kevinetal09} we obtain the ratio of the power spectrum as corrected with the fitted beam to the power spectrum as it should have been corrected by the true beam. 
In Figure~\ref{fig:ratiosk} we display lines which bound 68\% of the ensemble transfer functions for Planck channels.

The simulation of repeated Jupiter calibrations id done in such way that each template is an unbiased estimator of the true template. But in real life, they could be a biased estimator (for instance the Planck pointing error could bias the beam function always in the same direction). This prompted us to consider the runs presented in Section~\ref{wbeam}.

\begin{figure}[t]
\includegraphics[width=0.49\linewidth,height=0.3\linewidth]{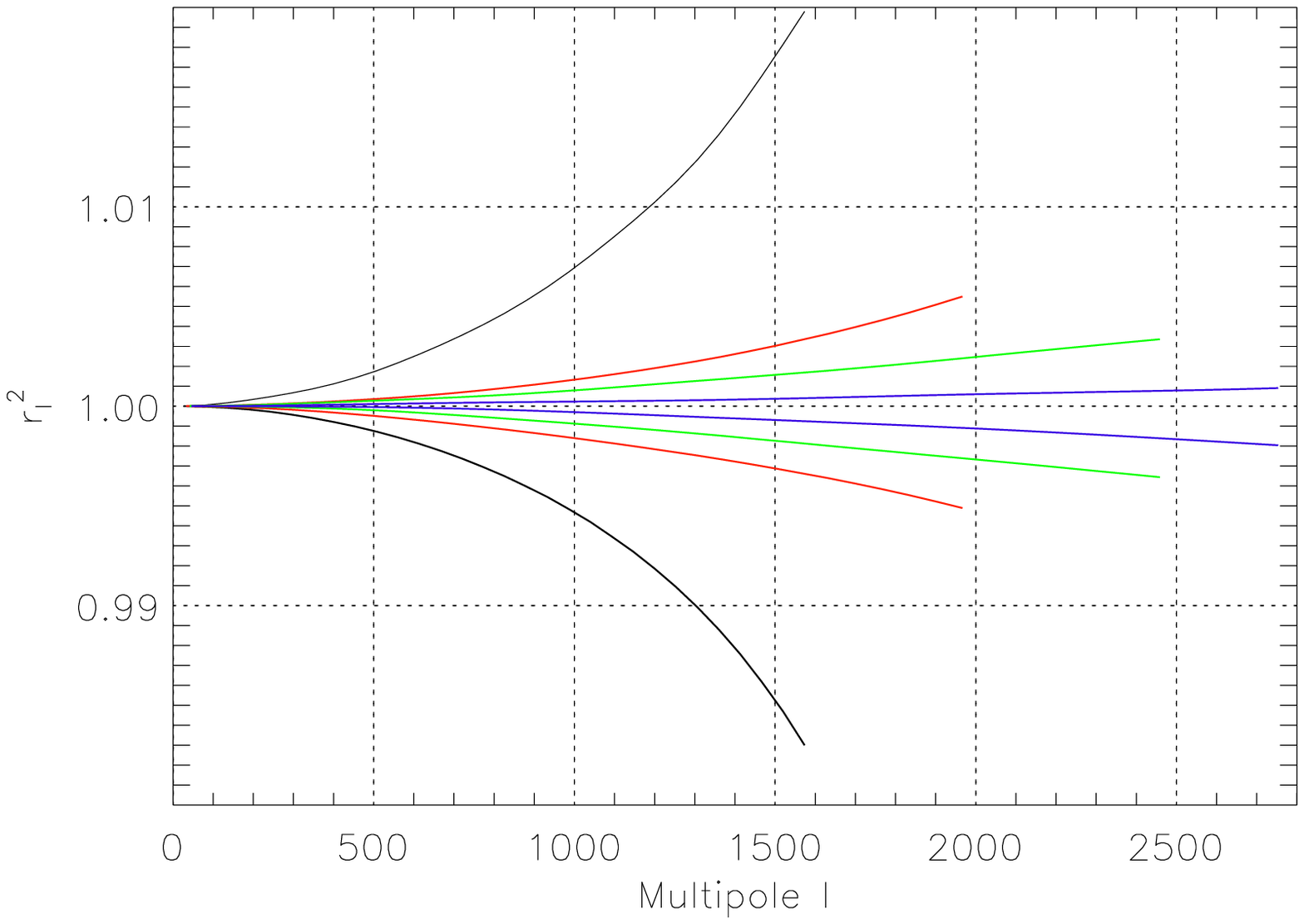}
\includegraphics[width=0.49\linewidth,height=0.3\linewidth]{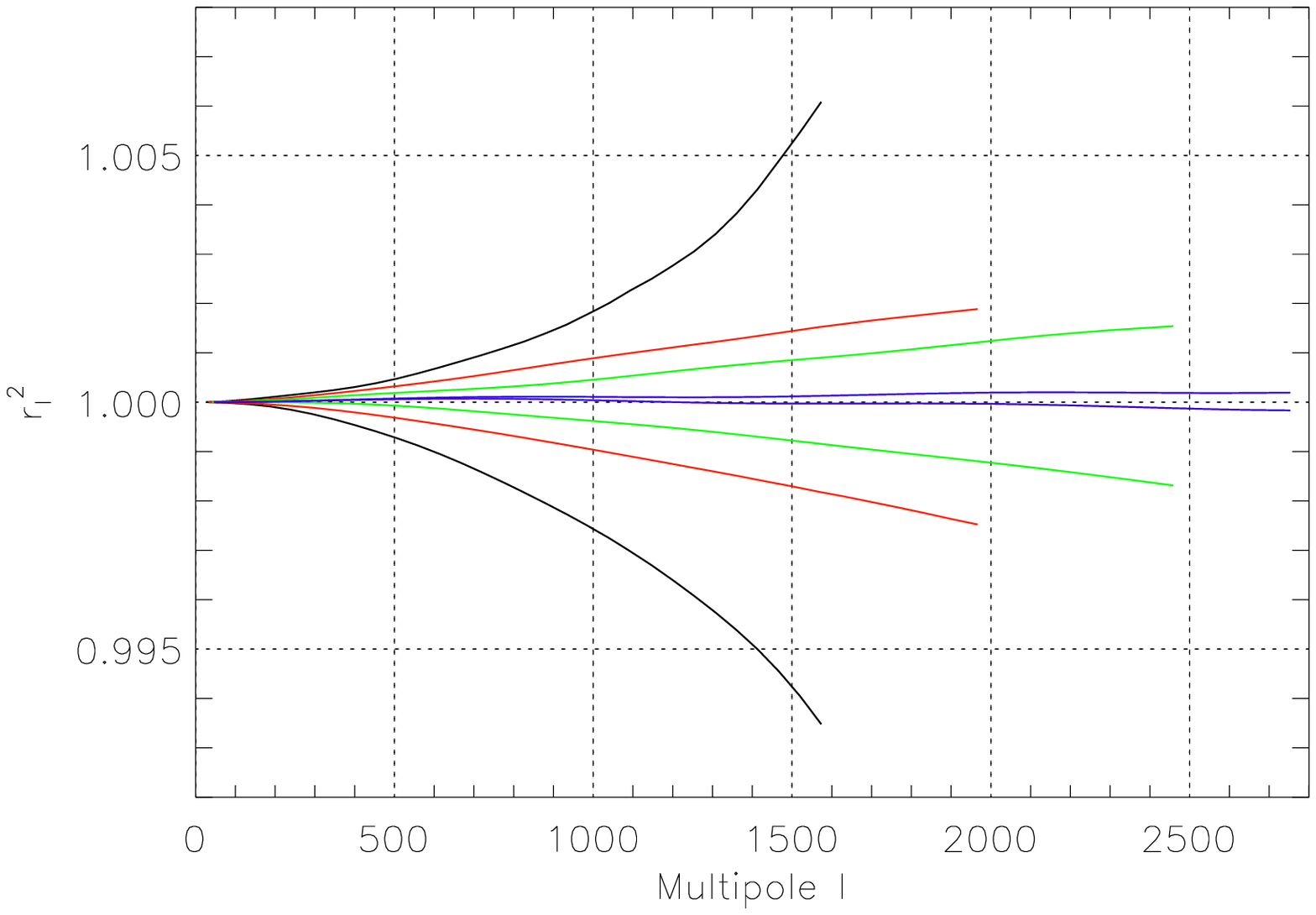}
\label{fig:ratios3}
\caption{Extreme (left) and Mild (right) beam transfer functions for the Planck 70\,GHz (black), 100\,GHz (red), 143\,GHz (green) and 217\,GHz (blue) channels obtained from beam fitting applied to  destriped data.\label{fig:ratiosds} }
\end{figure}

\begin{figure}[t]
\begin{center}
\includegraphics[width=0.49\linewidth,height=0.3\linewidth]{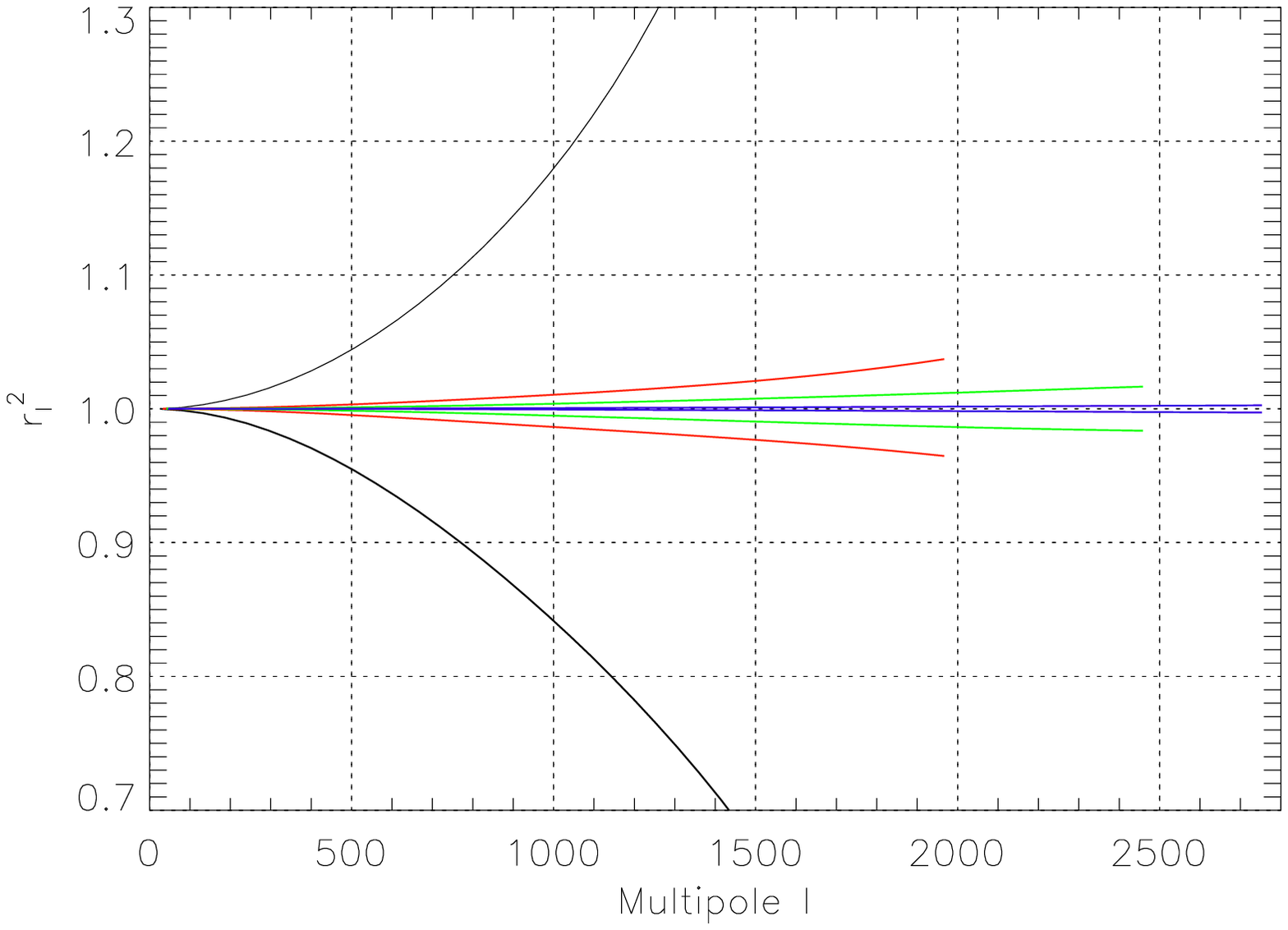}
\includegraphics[width=0.49\linewidth,height=0.3\linewidth]{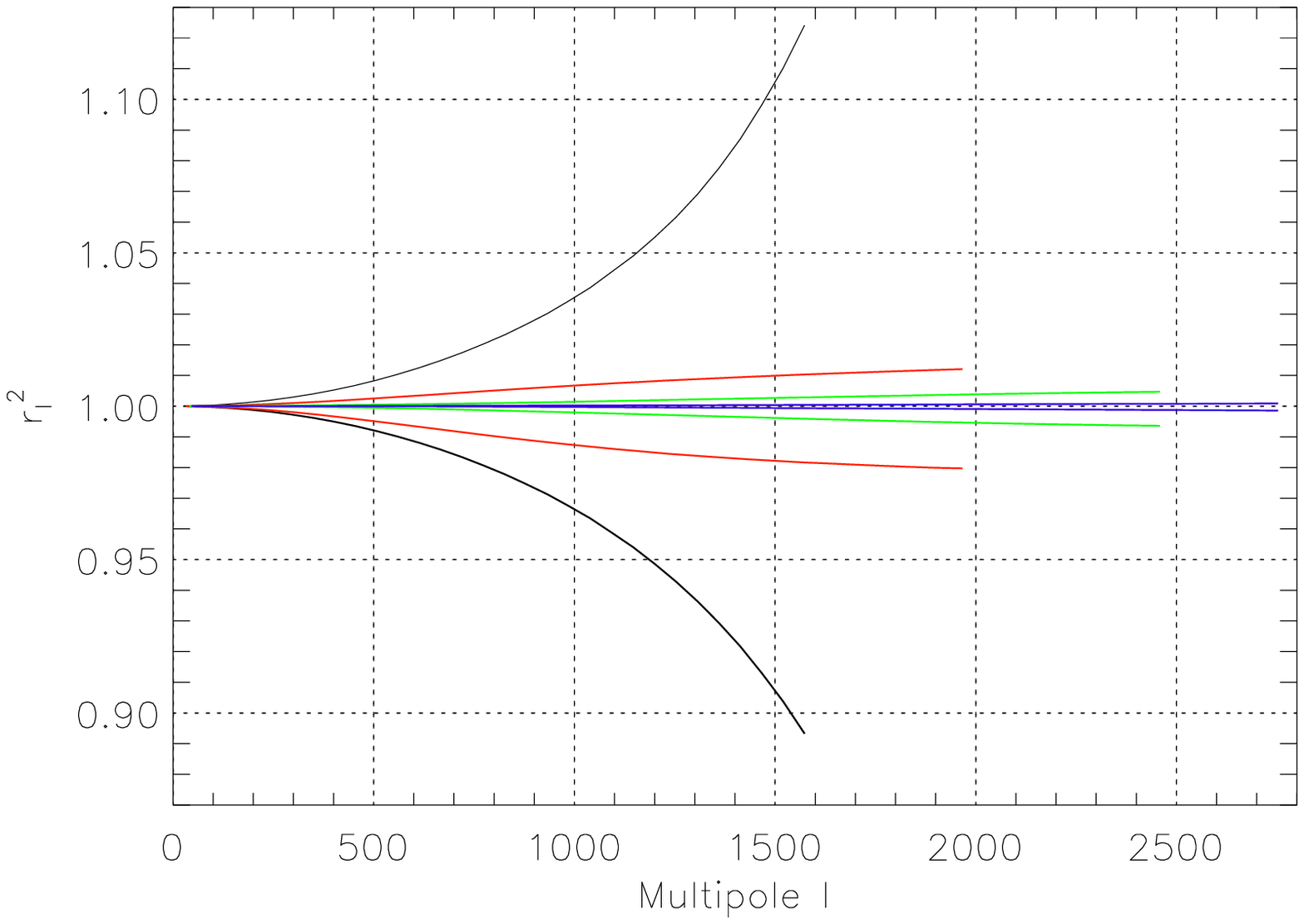}
\caption{Extreme (left) and Mild (right) beam transfer functions for the Planck 70\,GHz (black), 100\,GHz (red), 143\,GHz (green) and 217\,GHz (blue) channels obtained from beam fitting applied to data with white + $1/f$ noise.  \label{fig:ratios}}
\end{center}
\end{figure}

\begin{figure}[t]
\begin{center}
\includegraphics[width=0.45\linewidth]{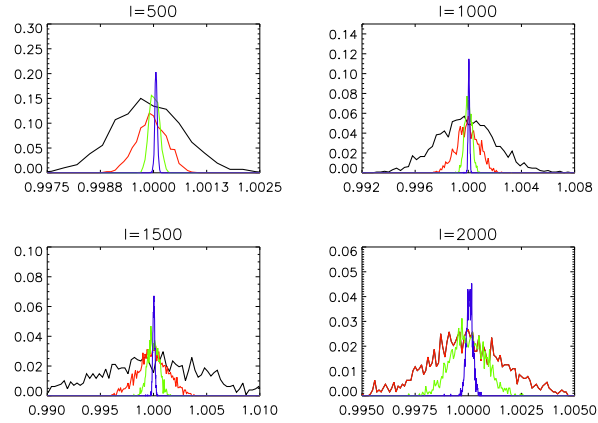}
\caption{Normalized distributions of the beam transfer functions, ${\cal{B}}_{\ell} = (B_{\ell}^{r})^{2}$ for multipoles $\ell = 500, 1000, 1500, 2000$ for 70\,GHz 
(black), 100\,GHz (red), 143\,GHz (green), and 217\,GHz(blue) obtained from the beam fitting on destriped data. \label{fig:ratioshist}}
\end{center}
\end{figure}

The Beam fitting is applied to data with white + $1/f$ noise, and to destriped data, i.e., after application e.g of a ``destriping'' mapmaking code which removes almost all of the effects of $1/f$ noise [\cite{Poutanen:2006, Ashdown:2007ta, Ashdown:2007tb, Ashdown:2009}].  We use realistic Grasp beams and the parametric model of the reconstructed beams (the results with non-parametric model will be presented in a future paper).
Figure~\ref{fig:ratiosds} shows extreme and mild beam transfer functions for the Planck 70\,GHz, 100\,GHz, 143\,GHz and 217\,GHz channels obtained from the beam fitting procedure applied to destriped data (hence containing a a very low level of $1/f$ residuals).
For comparison purposes we plot in Figure~\ref{fig:ratios} these functions obtained from data with a white and $1/f$ noise background.
We also plot in Figures~\ref{fig:ratioshist} the normalized histograms of the ratios, $r_{\ell}^{2}$,  for singe multipoles $\ell=500, 1000, 1500, 2000$. For all channels the distributions are slightly skewed and get broader with increasing multipole $\ell$ for each channel. 
We can also compare the probability of the mildly deviated and extremely deviated transfer functions used in \S\,~\ref{results}.
For instance for 70\,GHz for $\ell=1000$  the mild function is $\simeq 10 \%$ probable while the extreme function is approximately 100 times less likely.
The maximum variation for transfer function ratios, $r_{\ell}^{2}$, is of the order $2 \%$ for 70\,GHz ($0.5 \%$ for 100GHz) for destriped data, while for white and $1/f$ noise data with no attempt at destriping it increases to $\sim30 \%$ for the 70\,GHz channel ($\sim2.5 \%$ for 100\,GHz).

In \S\,~\ref{results} we infer that the parameter constraints from beams obtained with destriped data are slightly worse but very close to those obtained with a white noise background as expected.

\section{Analysis: from Beam transfer function uncertainties to parameter estimation \label{beam-analysis}}

To propagate the beam measurement errors to parameters we apply the \texttt{MCBR} method following the procedure described in \S\,~\ref{mcbr}. We make use of the beam transfer functions obtained with the beam fitting described in \S\,~\ref{beam-fitting}. For this purpose we use a modified version of  \texttt{cosmomc} with built-in \texttt{MCBR} step as described in \S\,~\ref{mcbr}. We consider a set of five chains.  The convergence
diagnostic is based on the Gelman and Rubin statistic, as usual in the field. 
Following \texttt{MCBR}, we choose randomly the beam transfer function (from the set of $1280$ simulations)  for each step of the Markov Chain. 
We sample a six-dimensional set of cosmological
parameters, with flat priors: the physical baryon and Cold Dark Matter densities,
$\omega_b=\Omega_bh^2$ and $\omega_c=\Omega_ch^2$; the ratio of the sound horizon to the angular diameter distance at decoupling, $\theta_s$; the scalar spectral index
$n_S$; the overall normalization of the spectrum $log[10^{10} A]$ at $k=0.05$
Mpc$^{-1}$ (hereafter $A_S$), and the optical depth to reionization $\tau$. 
We use a cosmic age top-hat prior 10\,Gyr$ \le t_0 \le$ 20\,Gyr,  consider purely adiabatic initial conditions only, we impose flatness, and we treat the dark energy component as a cosmological constant.
  
\begin{figure}[t]
\begin{center}
\includegraphics[width=0.45\linewidth]{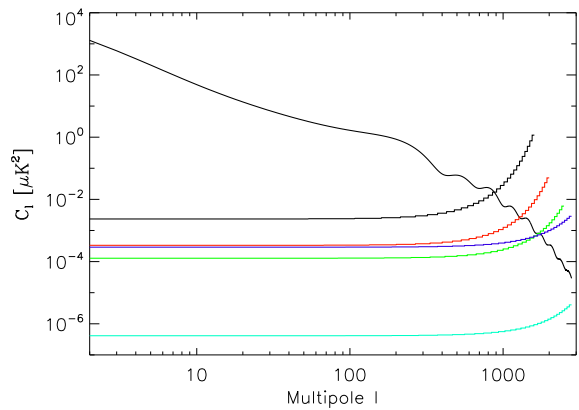}
\caption{CMB angular power spectrum (best fit of WMAP 1yr, black line) and noise levels for Planck: $70\,GHz$(black), $100\,GHz$(red), $143\,GHz$(green), $217\,GHz$(blue) and for Epic $150\,GHz$(cian).} \label{fig:models}
\end{center}
\end{figure}

We create simulated datasets with the noise properties of the Planck 70, 100, 143 and
217\,GHz (\cite{:2006uk}) channels, as well as one example of a future experiment. For the latter we considered the noise levels of Epic 150\,GHz (\cite{Bock:2008ww}).  We take as our cosmological model  the best fit of WMAP 1yr: $\Omega_bh^2=0.02238$; $\Omega_ch^2= 0.11061$; $H_0 = 71.992$; $\tau=0.110267$; $n_S =0.95820$; and $A_S = 3.0824$ (\cite{Spergel:2003cb}).
These simulated datasets are given in terms of the angular Power Spectrum $C^{obs}_{\ell}$ as described in \S\,~\ref{mcbr}.
We compute the noise ${\cal{N}}_{\ell}=(\Delta T \times fwhm)^2$ for Planck and Epic from the sensitivity $\Delta T/T$ and 
the nominal $fwhm$ of the beam assuming a Gaussian profile (tabulated in Table~\ref{tab:exp}).
In Figure~\ref{fig:models}  we plot the theoretical model vs.~the noise levels for each channel considered. Results from this analysis are given in \S\,~\ref{results}.

\begin{table}[htb]\footnotesize
\begin{center}
\begin{tabular}{rcccc}
Experiment & Channel & FWHM & $\Delta T/T$ & FWHM (grasp)\\
\hline
Planck
& 70  & $14'$ & 4.7 & 13' \\
& 100 & $10'$ & 2.5  & 9.22'\\
& 143 & $7.1'$ & 2.2 & 6.49'\\
& 217 & $5.0'$ & 4.8 & 4.48'\\
\hline
Epic-CS & 150  & $5.0'$ & 0.81 \\

\end{tabular}
\caption{Planck (\cite{:2006uk}) and Epic (\cite{Bock:2008ww}) experimental specifications.  Channel frequency is given
in GHz, FWHM in arcminutes and noise in $10^{-6}$. The last column gives the fwhm of our fiducial beams based on a Gaussian fit to the realistic GRASP beams.}
\label{tab:exp}
\end{center}
\end{table}

\section{Results \label{results}}

\subsection{Results: effect of beam uncertainties \label{uncbeam}}

\begin{figure*}
\begin{center}
\includegraphics[scale=0.75]{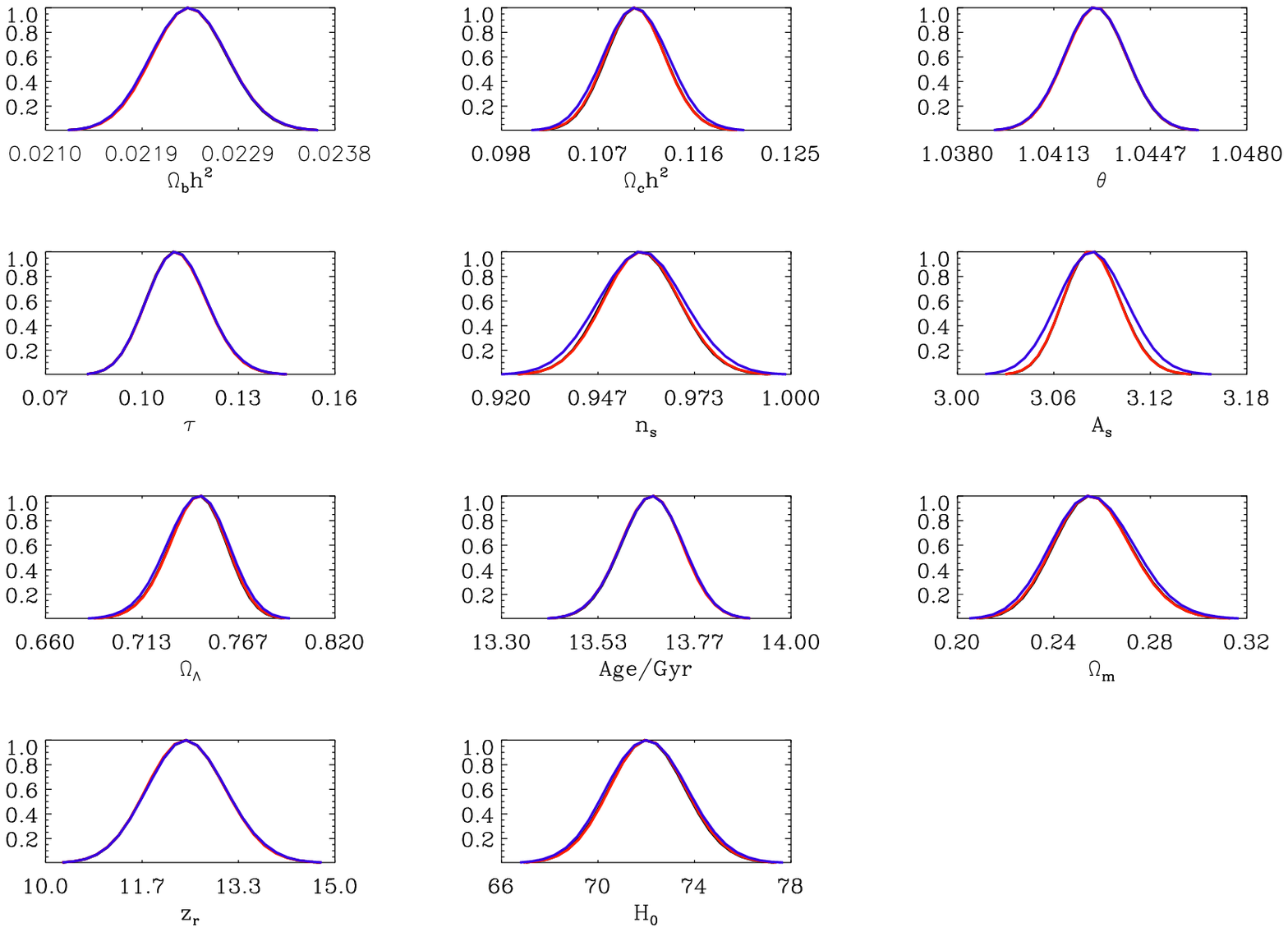}
\caption{Marginalized parameter constraints for Planck 70 GHz without beam uncertainty (black), marginalized over the beam uncertainty via \texttt{MCBR} considering the destriped data (red), and in the presence of white noise $+ 1/f$ noise (blue).} \label{fig:p70_compds}
\end{center}
\end{figure*}

\begin{figure*}
\begin{center}
\includegraphics[scale=0.75]{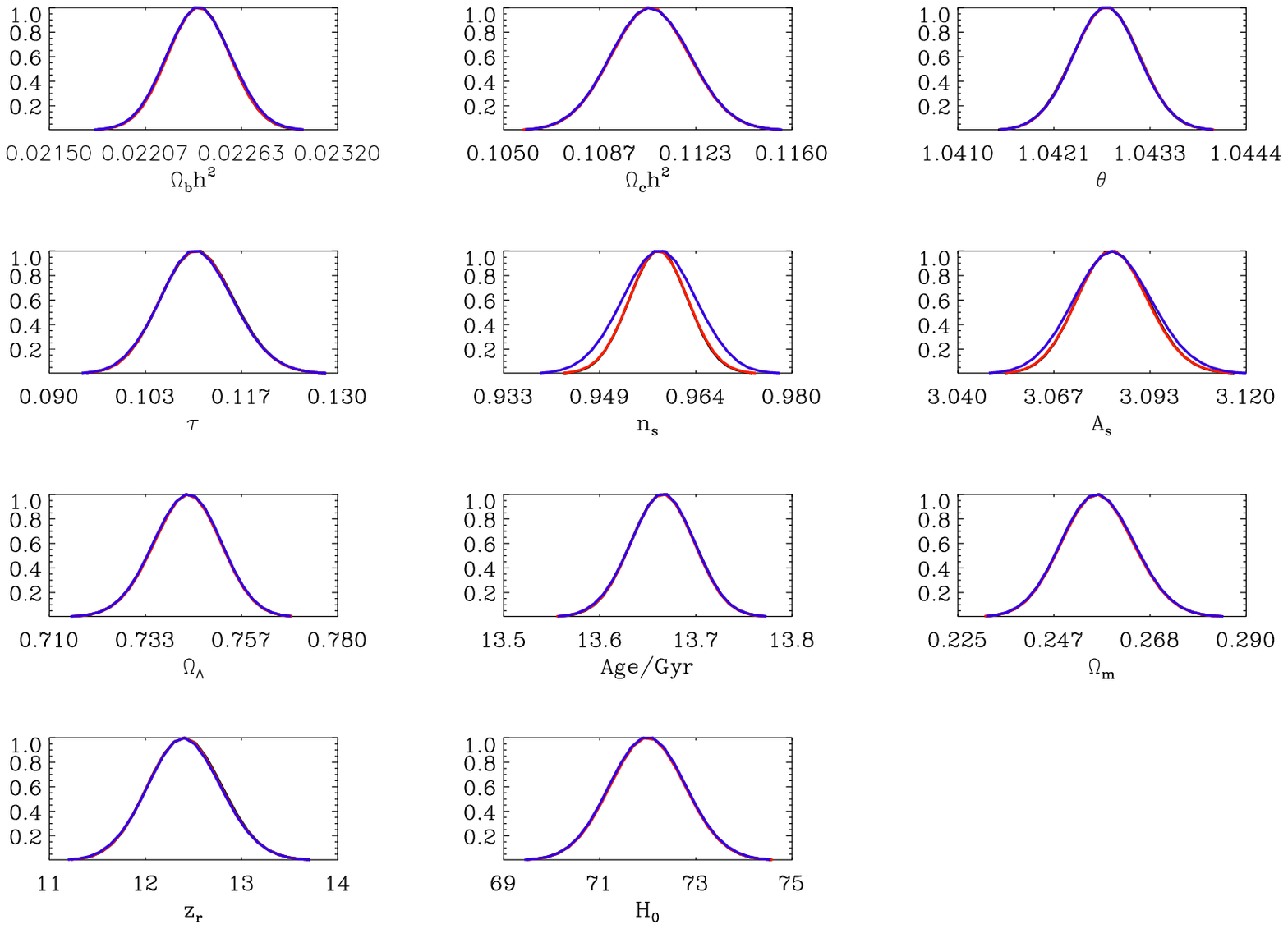}
\caption{Marginalized parameter constraints for Planck 100 GHz without beam uncertainty (black), marginalized over the beam uncertainty via \texttt{MCBR} considering
 the destriped data(red) and in the presence of white noise $+ 1/f$ noise (blue).} \label{fig:p100_compds}
\end{center}
\end{figure*}

\begin{figure*}
\begin{center}
\includegraphics[scale=0.75]{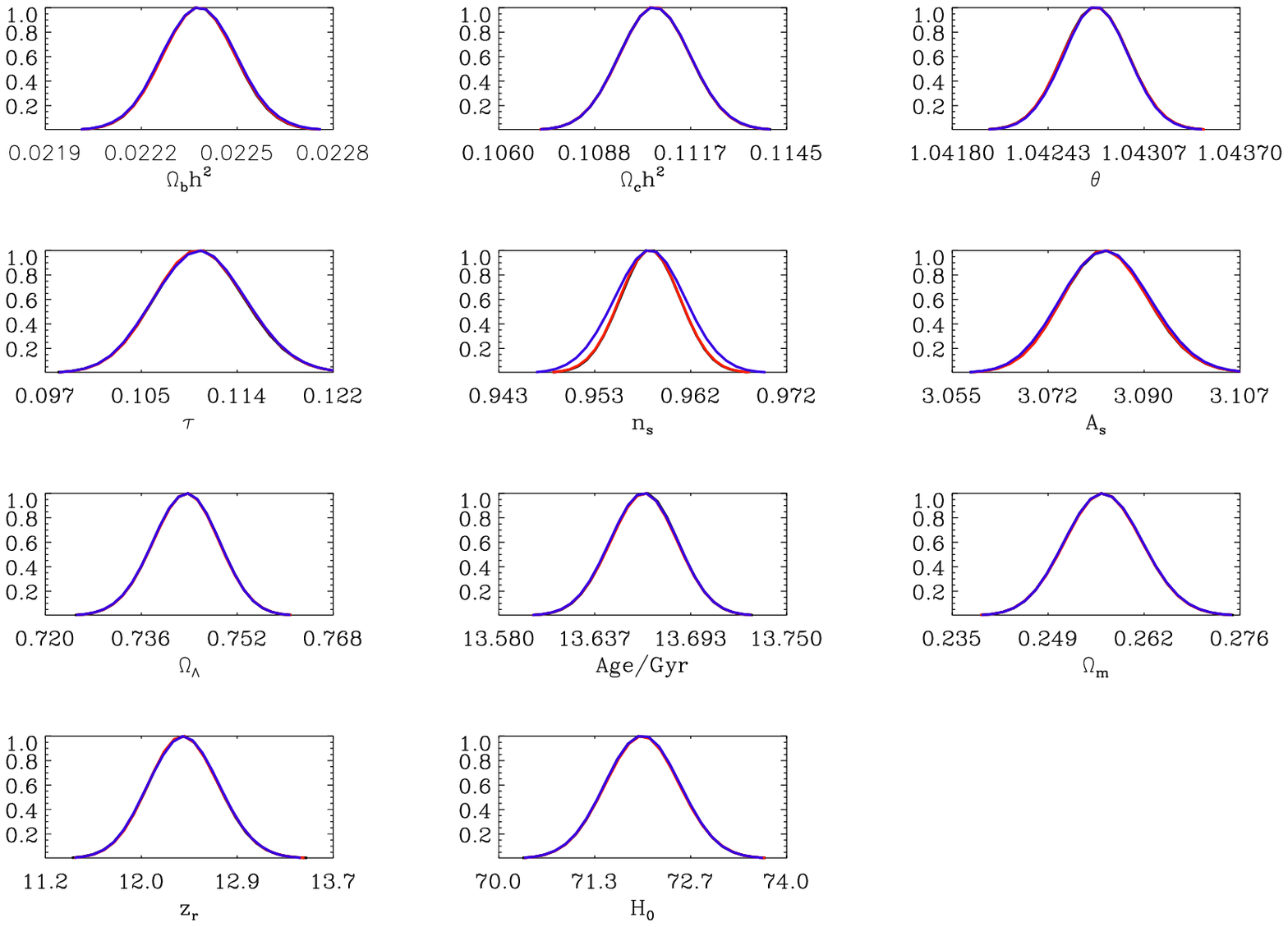}
\caption{Marginalized parameter constraints for Planck 143 GHz without beam uncertainty (black), marginalized over the beam uncertainty \texttt{MCBR} considering the destriped data (red) and in the presence of white noise $+ 1/f$ noise (blue).} \label{fig:p143_compds}
\end{center}
\end{figure*}

\begin{figure*}
\begin{center}
\includegraphics[scale=0.75]{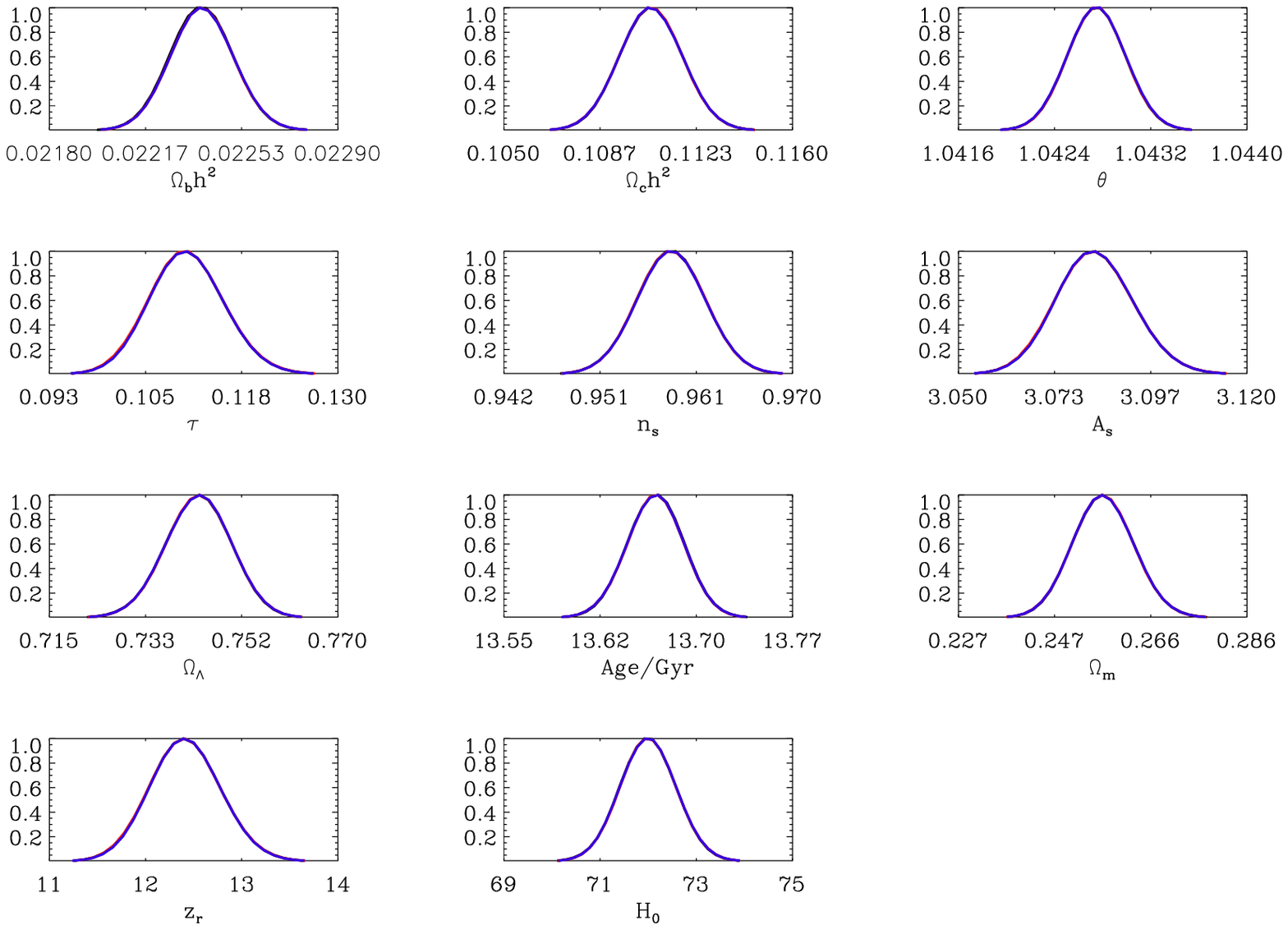}
\caption{Marginalized parameter constraints for Planck 217 GHz without beam uncertainty (black), marginalized over the beam uncertainty \texttt{MCBR} considering the destriped data (red) and in the presence of white noise $+ 1/f$ noise (blue).} \label{fig:p217_compds}
\end{center}
\end{figure*}

\begin{figure*}
\begin{center}
\includegraphics[width=12cm, height=5cm]{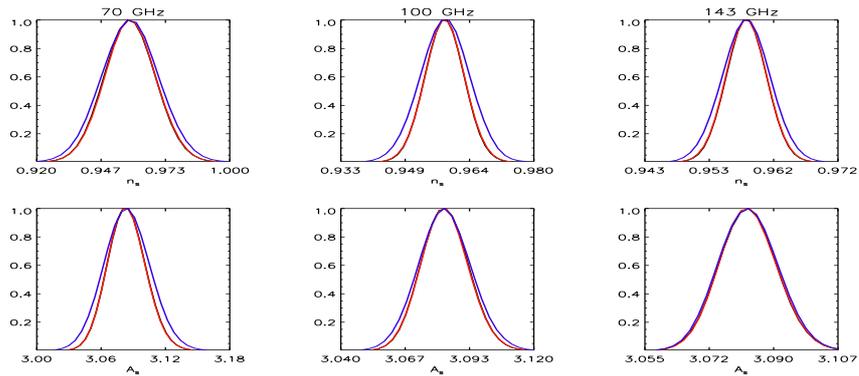}
\caption{Marginalized constraints for  the most impacted parameters, $n_{s}$ and $A_{s}$,  for Planck channels 70GHz, 100GHz and 143GHz,  without beam uncertainty (black), marginalized over the beam uncertainty considering 
the destriped data (red) and in the presence of white noise $+ 1/f$ noise (blue).} \label{fig:all_compds}
\end{center}
\end{figure*}

Figures~\ref{fig:p70_compds}, \ref{fig:p100_compds}, \ref{fig:p143_compds}, \ref{fig:p217_compds} show the marginalized parameter constraints for Planck in three cases: without beam uncertainty (i.e., considering the true fiducial beam) (black); with beam uncertainty using the beam transfer functions obtained using the destriped data (red); and in the presence of $1/f$ noise (blue).
Table~\ref{tab:resultsall} gives the input cosmological parameters and the mean values and marginalized $68\%$ confidence limits obtained after accounting for the beam errors.  To facilitate comparisons, Figure~\ref{fig:all_compds} shows these same marginalized constraints for $n_{s}$ and $A_{s}$, the parameters where the largest differences are seen between the three cases.  
 
\begin{figure*}
\begin{center}
\includegraphics[scale=0.75]{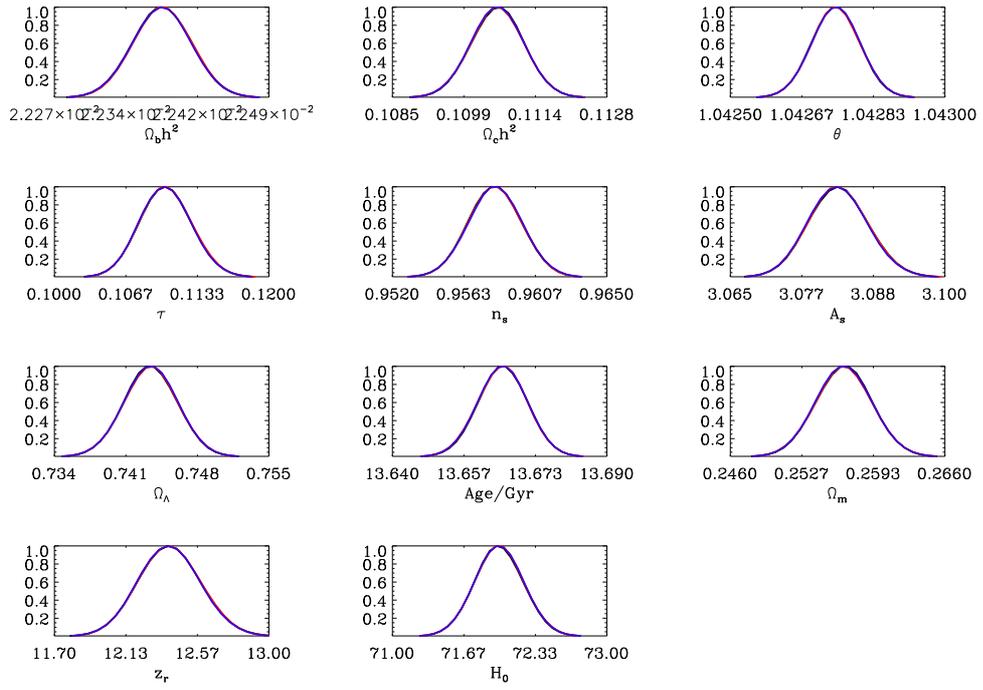}
\caption{Marginalized parameter constraints for a future experiment with Epic 150 GHz specifications without beam uncertainty (black), marginalized over the beam uncertainty considering 
the destriped data (red) and in the presence of white noise $+ 1/f$ noise (blue).} \label{fig:e150_compds}
\end{center}
\end{figure*}

Equivalent results for a more sensitive polarization experiment---Epic 150\,GHz---are plotted in Figure~\ref{fig:e150_compds}; corresponding parameter values are given in table~\ref{tab:resultsall}. 

The most noticeable effect of beam uncertainties is to widen the marginal distributions of some parameters, especially $n_s$, for uncertainties obtained in the presence of $1/f$ noise but without destriping.  $A_s$, and $\Omega_bh^2$ are also affected.
In this case the distribution of the fitted beam transfer functions is wider than that obtained from white noise or destriped data as shown in Figures~\ref{fig:ratiosds} and \ref{fig:ratios}. 

\begin{figure}[t]
\begin{center}
\includegraphics[width=0.45\linewidth]{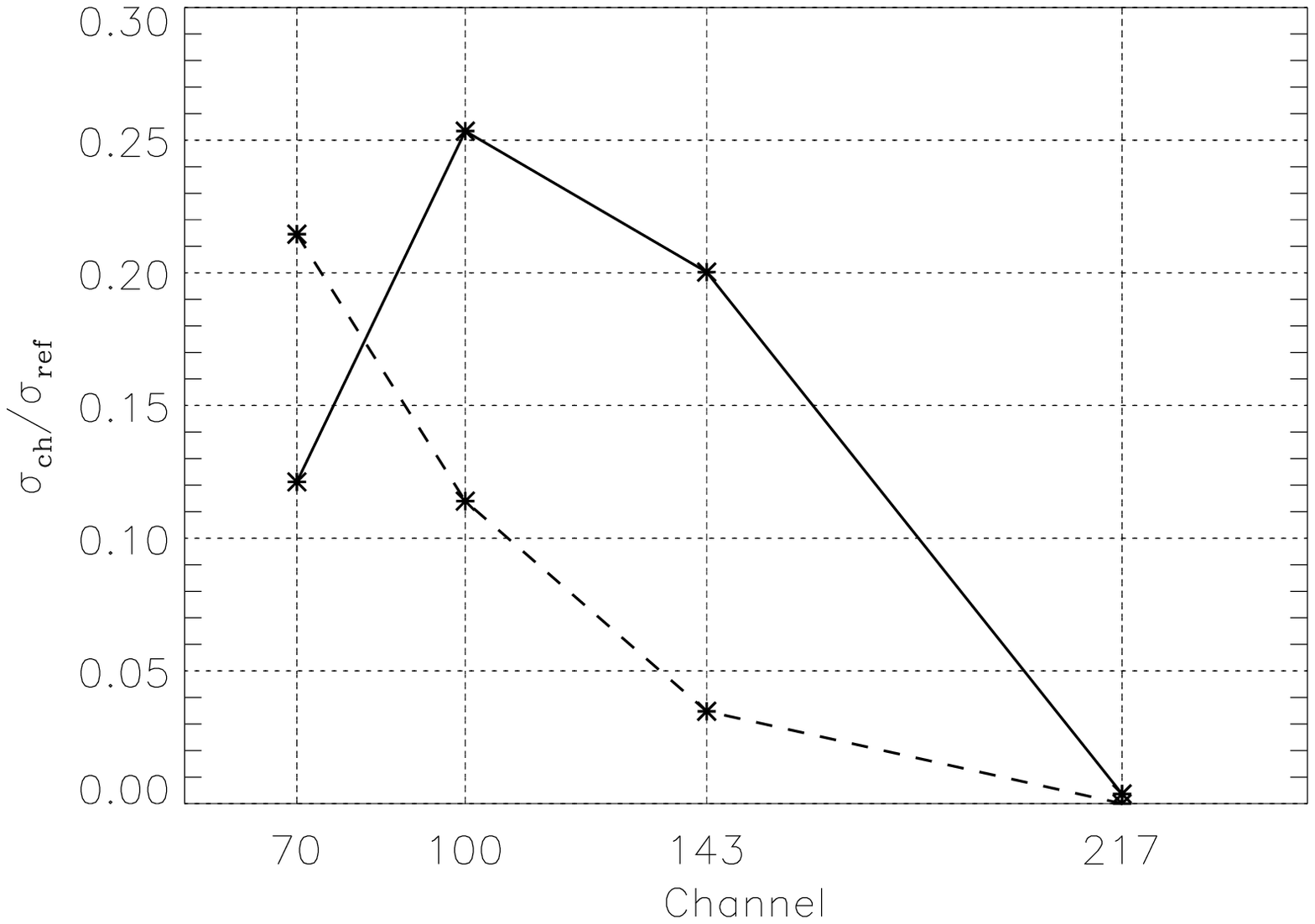}
\caption{Enhancement factor, $\sigma_{ch}/\sigma_{ref}$ for $n_{s}$ (solid line) and $A_{s}$ (dashed line), where $\sigma_{ch}$ is the width of the distribution when beam errors are marginalized over by applying \texttt{MCBR}; $\sigma_{ref}$ is the width of the distribution for the simulated data convolved with the fiducial beam (no  beams errors included), for beams fitted on data with white and $1/f$ noise background.}  \label{fig:sigmafrac}
\end{center}
\end{figure}

Define $\sigma_{ch}$, the width of the distribution when beam errors are marginalised by applying \texttt{MCBR}, and $\sigma_{ref}$, the width of the distribution for the simulated data convolved with the fiducial beam (with no beam errors included).
Figure~\ref{fig:sigmafrac} shows the enhancement factor, $\sigma_{ch}/\sigma_{ref}$, for parameters $n_{s}$ and $A_{s}$ for beams fitted on data with white + $1/f$ noise.  For example, at  100\,GHz the distributions of $n_{s}$ and $A_{s}$ widen by 25\% and 11\%, respectively.

This widening is much reduced by the use of destriping techniques.  For example, with destriping the uncertainties in the beams are 0.5\% for 100\,GHz at $\ell=1500$, which translates into an increase of parameter uncertainties of 0.1\%.  Without destriping, the uncertainties on $A_{s}$  at 70\,GHz and on $n_s$ at 100\,GHz increase by 21\% and 25\%, respectively, for beams fitted on white + $1/f$ noise data.  This is a convincing demonstration of the relevance and power of destriping techniques in reducing the effect of $1/f$ noise for Planck.

\begin{table*}                                                                   
\begin{center}                                                                   
\begin{tabular}{llccc}                                                           
Channel&Parameter &  no beam uncertainty & destriped & white noise$+1/f$\\                   
\hline                                                                           
\hline                                                                           
Planck 70 GHz&$\Omega_bh^2$ &$0.22393\pm0.00035$& $0.22401\pm0.00035$  & $0.22394\pm0.00036$ \\
&$\Omega_ch^2$ &$0.1106\pm0.0027$ & $0.1105\pm0.0027$ & $0.1106\pm0.0029$ \\                   
&$\theta $ &  $1.0428\pm0.0010$ & $1.0428\pm0.0010$ & $1.0428\pm0.0010$ \\                     
&$\tau$ & $0.1112\pm0.0091$ & $0.1111\pm0.0091$ & $0.1112\pm0.0091$ \\
&$n_s$ & $0.959 \pm 0.010$ & $0.959\pm0.010$ & $0.959\pm0.011$ \\
&$A_s$ & $3.084 \pm 0.017$ & $3.084\pm0.017$ & $3.084\pm0.021$ \\
\hline
\hline
Planck 100 GHz&$\Omega_bh^2$ &$0.22383\pm0.00018$& $0.22383\pm0.00018$  & $0.22383\pm0.00018$ \\
&$\Omega_ch^2$ &$0.1106\pm0.0015$ & $0.1106\pm0.0015$ & $0.1106\pm0.0015$ \\
&$\theta $ &  $1.04275\pm0.00038$ & $1.04275\pm0.00038$ & $1.04275\pm0.00038$ \\
&$\tau$ & $0.1107\pm0.0049$ & $0.1106\pm0.0049$ & $0.1106\pm0.0049$ \\
&$n_s$ & $0.9583 \pm 0.0046$ & $0.9583\pm0.0046$ & $0.9583\pm0.0058$ \\
&$A_s$ & $3.0832 \pm 0.0094$ & $3.0831\pm0.0094$ & $3.083\pm0.011$ \\
\hline
\hline
Planck 143 GHz&$\Omega_bh^2$ &$0.22381\pm0.00011$& $0.22381\pm0.00011$  & $0.22381\pm0.00011$ \\
&$\Omega_ch^2$ &$0.1106\pm0.0010$ & $0.1106\pm0.0010$ & $0.1106\pm0.0010$ \\
&$\theta $ &  $1.04275\pm0.00021$ & $1.04275\pm0.00021$ & $1.04274\pm0.00021$ \\
&$\tau$ & $0.1105\pm0.0038$ & $0.1105\pm0.0038$ & $0.1105\pm0.0039$ \\
&$n_s$ & $0.9582 \pm 0.0029$ & $0.9582\pm0.0029$ & $0.9582\pm0.0035$ \\
&$A_s$ & $3.0829 \pm 0.0074$ & $3.0828\pm0.0075$ & $3.0829\pm0.0077$ \\
\hline
\hline
Planck 217 GHz&$\Omega_bh^2$ &$0.22380\pm0.00012$& $0.22382\pm0.00012$  & $0.22383\pm0.00012$ \\
&$\Omega_ch^2$ &$0.1106\pm0.0012$ & $0.1106\pm0.0012$ & $0.1106\pm0.0012$ \\
&$\theta $ &  $1.04275\pm0.00023$ & $1.04275\pm0.00023$ & $1.04275\pm0.00024$ \\
&$\tau$ & $0.1106\pm0.0046$ & $0.1106\pm0.0046$ & $0.1107\pm0.0047$ \\
&$n_s$ & $0.9582 \pm 0.0032$ & $0.9583\pm0.0032$ & $0.9583\pm0.0032$ \\
&$A_s$ & $3.0831 \pm 0.0090$ & $3.0830\pm0.0090$ & $3.0831\pm0.0090$ \\
\hline
\hline
Epic 150 GHz&$\Omega_bh^2$ &$0.223802\pm0.000029$& $0.223798\pm0.000029$  & $0.223802\pm0.000029$ \\
&$\Omega_ch^2$ &$0.11061\pm0.00051$ & $0.11061\pm0.00051$ & $0.11061\pm0.00052$ \\
&$\theta $ &  $1.042750\pm0.000054$ & $1.042750\pm0.000054$ & $1.042750\pm0.00054$ \\
&$\tau$ & $0.1104\pm0.0023$ & $0.1103\pm0.0023$ & $0.1103\pm0.0024$ \\
&$n_s$ & $0.9582 \pm 0.0016$ & $0.9582\pm0.0016$ & $0.9583\pm0.0016$ \\
&$A_s$ & $3.0827 \pm 0.0047$ & $3.0825\pm0.0047$ & $3.083\pm0.0047$ \\
\end{tabular}
\caption{Mean values and marginalized $68\%$ c.l. limits using the fiducial beam: analysis without beam uncertainty (column 3), accounting the beam uncertainty from destriped data (column 4) and from the data with white and $1/f$ noise (column 5).}
\label{tab:resultsall}
\end{center}
\end{table*}

%

\subsection{Results: effect of assuming a wrong fiducial beam \label{wbeam}}

To illustrate the effect of incorrect beam assumptions we calculated parameters assuming a mildly and then an extremely `wrong' beam.  Specifically, we generated three simulated datasets, using: ${\cal{B}}_{\ell}$; a mildly wrong beam ${\cal{B}}_{\ell}^{mild}$;  and an extremely wrong beam ${\cal{B}}_{\ell}^{ext}$.  We analyzed these datasets with the modified version of \texttt{cosmomc} with  \texttt{MCBR} built in, and compared the cosmological parameters from the run with for the `true' fiducial beam ${\cal{B}}_{\ell}$ to those of both the mild and extreme deviated beams

\begin{figure*}
\begin{center}
\includegraphics[scale=0.75]{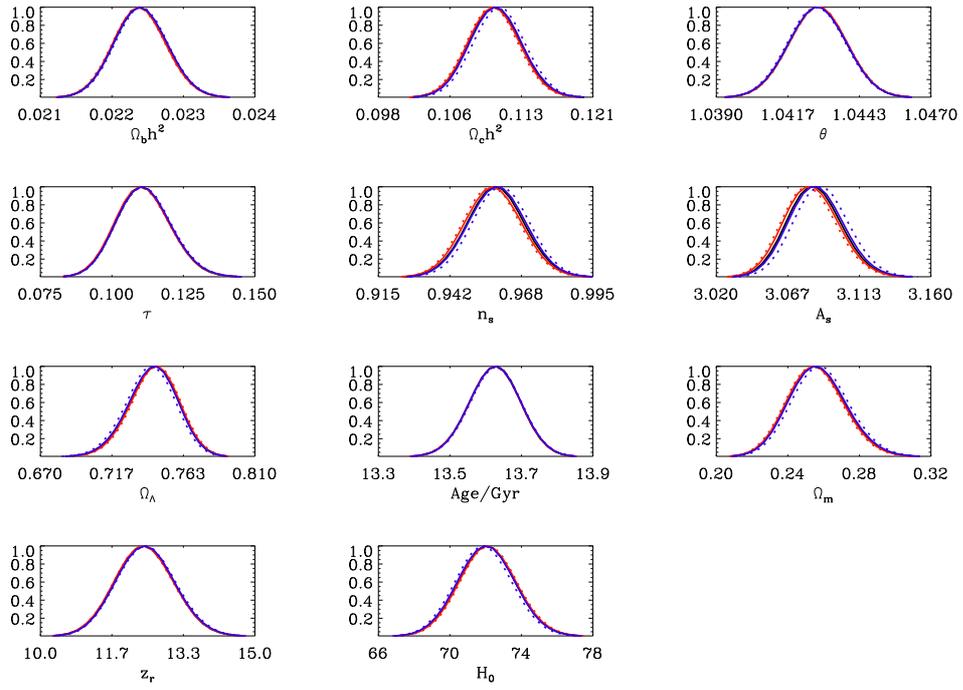}
\caption{Marginalized parameter constraints for Planck 70 GHz with beam randomization \texttt{MCBR}: true beam (black), decreasing function for destriped data (red),  increasing function for destriped data  (blue ), mild deviation (solid line) and extreme deviation from the true beam (dotted line)} \label{fig:p70_compfuncds}
\end{center}
\end{figure*}

\begin{figure*}
\begin{center}
\includegraphics[scale=0.75]{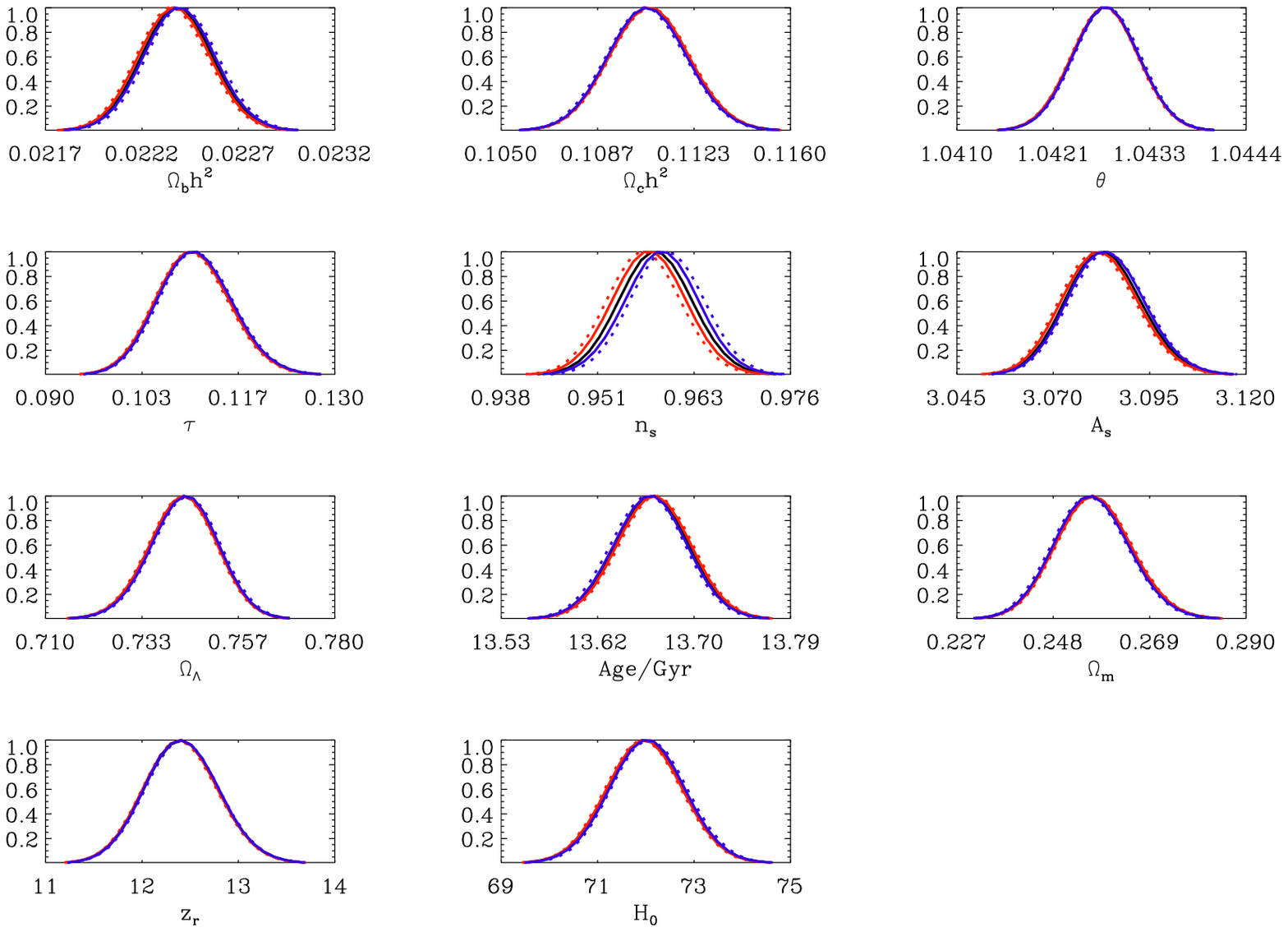}
\caption{Marginalized parameter constraints for Planck 100 GHz with beam randomization \texttt{MCBR}: true beam (black), decreasing function for destriped data (red),  increasing function for destriped data (blue ), mild deviation (solid line) and extreme deviation from the true beam (dotted line)} \label{fig:p100_compfuncds}
\end{center}
\end{figure*}

\begin{figure*}
\begin{center}
\includegraphics[scale=0.75]{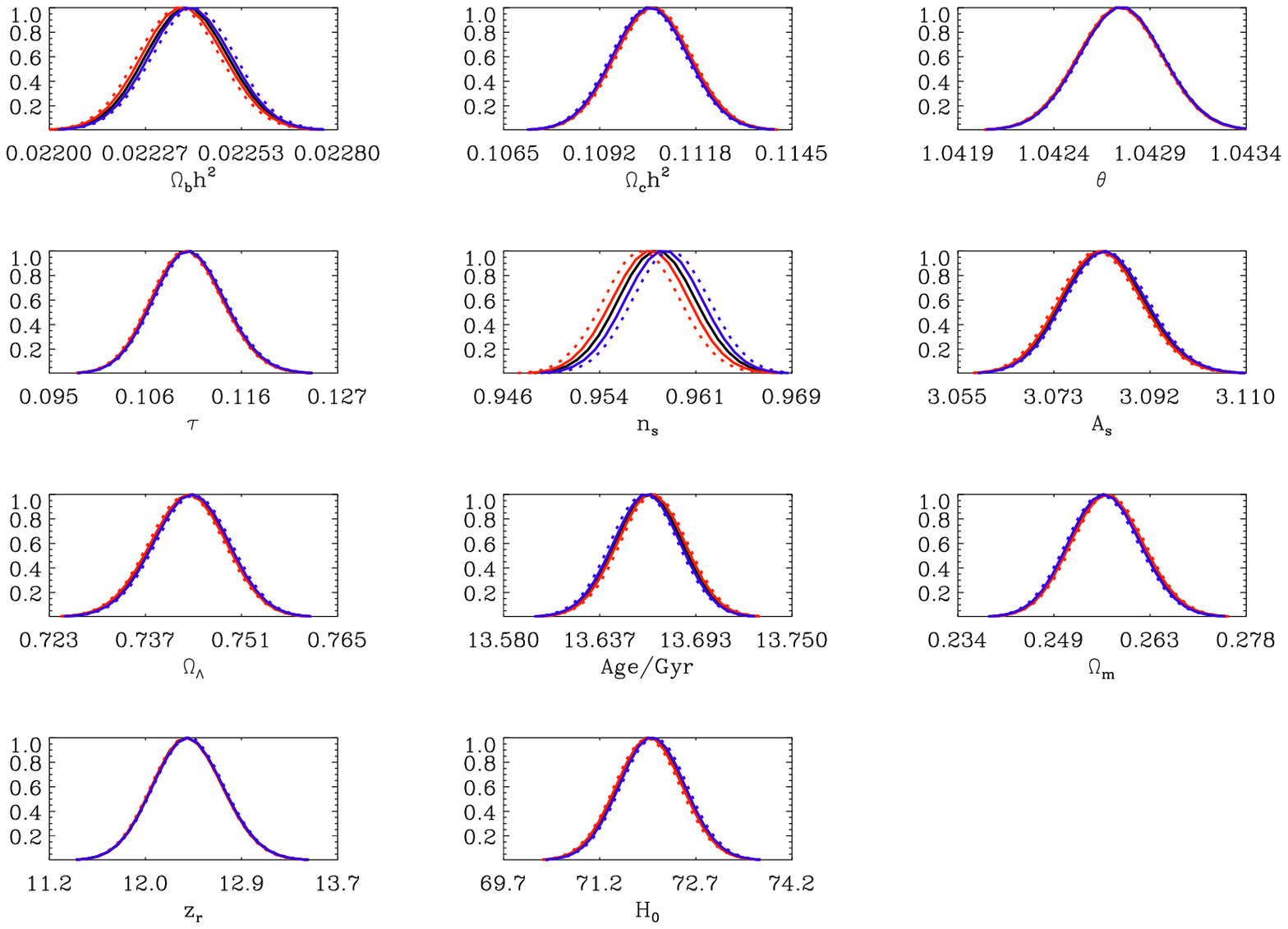}
\caption{Marginalized parameter constraints for Planck 143 GHz with beam randomization \texttt{MCBR} : true beam (black), decreasing function for destriped data (red),  increasing function for destriped data (blue ), mild deviation (solid line) and extreme deviation from the true beam (dotted line)} \label{fig:p143_compfuncds}
\end{center}
\end{figure*}

\begin{figure*}
\begin{center}
\includegraphics[scale=0.75]{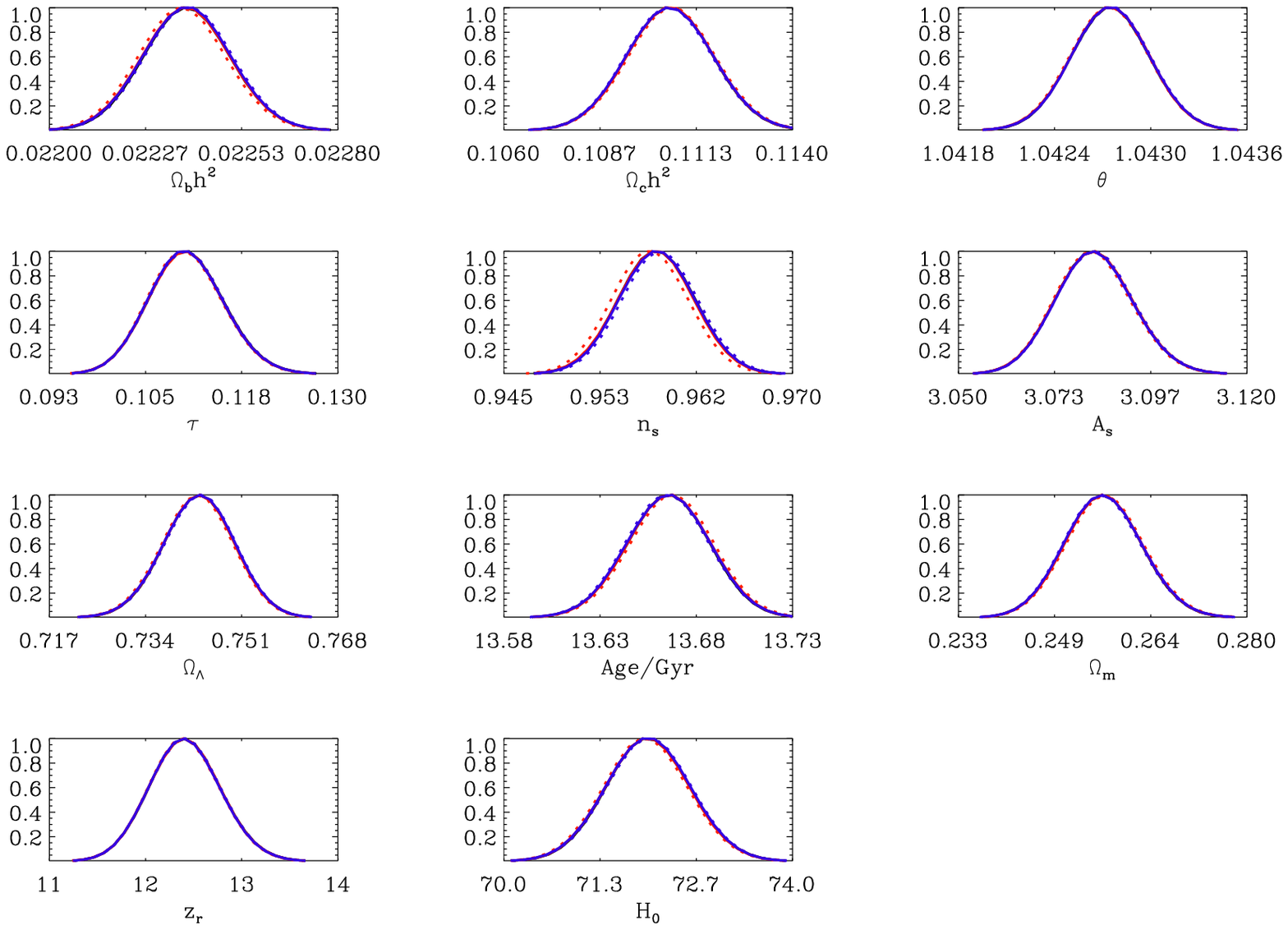}
\caption{Marginalized parameter constraints for Planck 217 GHz with beam randomization \texttt{MCBR}: true beam (black), decreasing function for destriped data (red),  increasing function for destriped data (blue ), mild deviation (solid line) and extreme deviation from the true beam (dotted line)} \label{fig:p217_compfuncds}
\end{center}
\end{figure*}

Figures~\ref{fig:p70_compfuncds},\ref{fig:p100_compfuncds},\ref{fig:p143_compfuncds},\ref{fig:p217_compfuncds} show marginalized parameter constraints from 70\,GHz, 100\,GHz, 143\,GHz,  and 217\,GHz, respectively, on destriped data.  We see that assuming an extreme beam deviation in the simulated data results in a biased estimation of some parameters, particularly $n_s$.  This is mostly due to incomplete marginalization, as we do not encompass an adequate distribution of deviations from the chosen fitted transfer function. 

\begin{figure*}
\begin{center}
\includegraphics[scale=0.75]{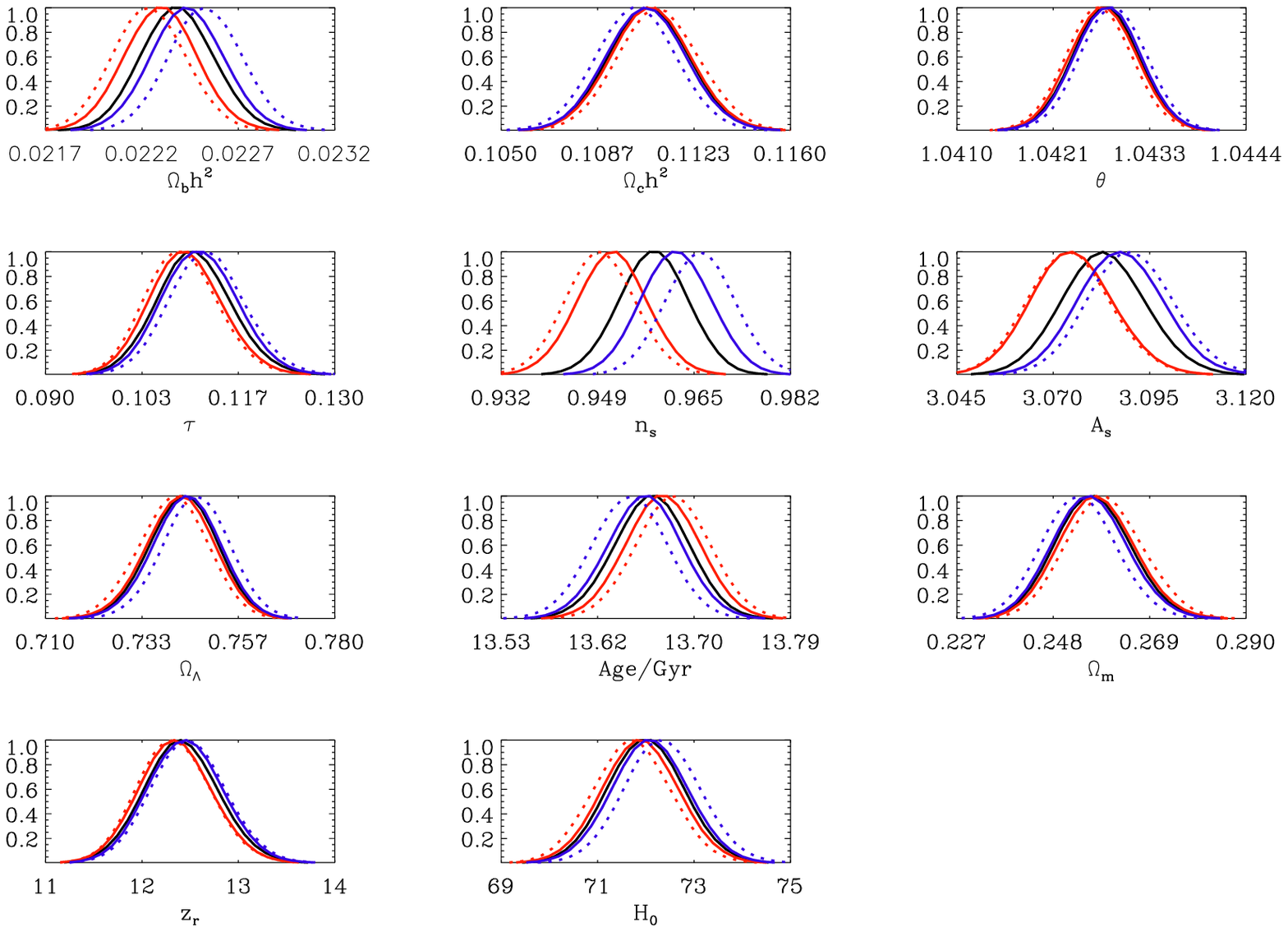}
\caption{Marginalized parameter constraints for Planck 100 GHz with beam randomization \texttt{MCBR}: true beam (black), decreasing function for white +1/f noise (red),  increasing function for white +1/f noise (blue ), mild deviation (solid line) and extreme deviation from the true beam (dotted line)} \label{fig:p100_compfunc}
\end{center}
\end{figure*}

\begin{figure*}
\begin{center}
\includegraphics[scale=0.75]{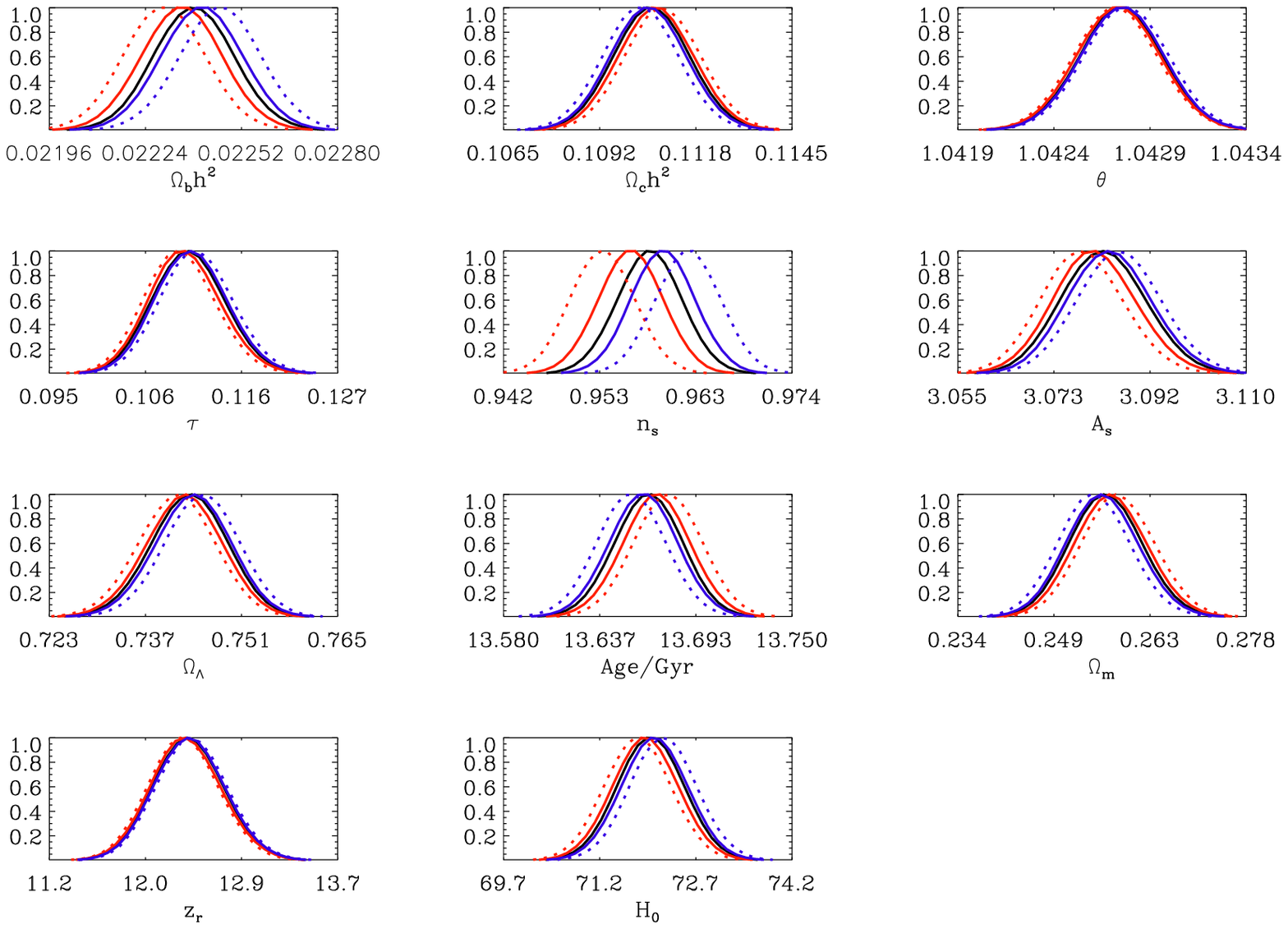}
\caption{Marginalized parameter constraints for Planck 143 GHz with beam randomization \texttt{MCBR}: true beam (black), decreasing function for white +1/f noise (red),  increasing function for white +1/f noise (blue ), mild deviation (solid line) and extreme deviation from the true beam (dotted line)} \label{fig:p143_compfunc}
\end{center}
\end{figure*}

For comparison, Figures~\ref{fig:p100_compfunc} and \ref{fig:p143_compfunc} show marginalized parameter constraints for the 100\,GHz and 143\,GHz channels, respectively, on data that have not been destriped.

\begin{figure}[t]
\begin{center}
\includegraphics[width=0.45\linewidth]{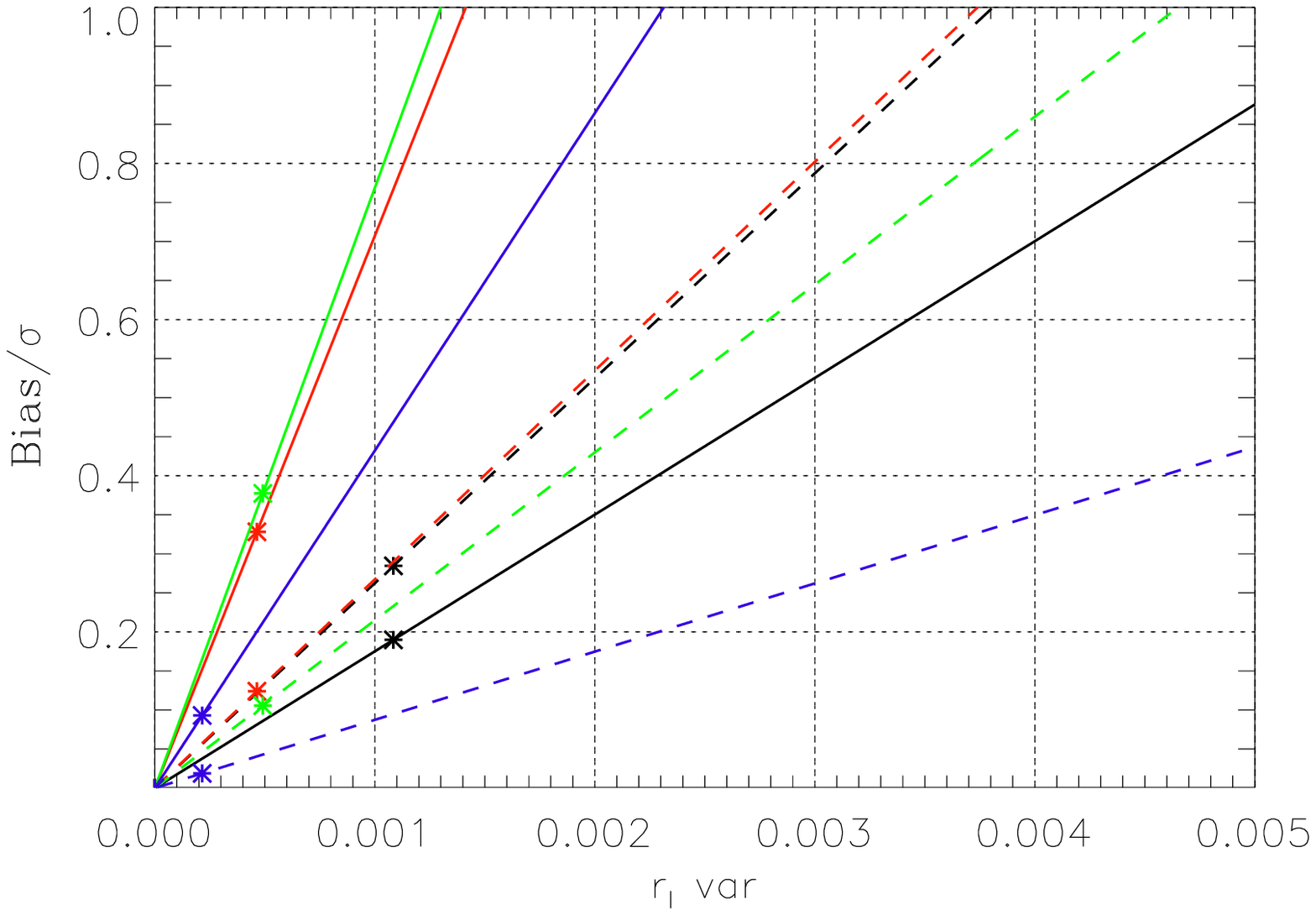}
\caption{Bias on $n_{s}$(solid line) and $A_{s}$(dotted line) in units of the error for the extreme beam functions, $r^{ext}_{\ell}$  for $\ell = 1/\sigma_{\nu}$ after beam randomization \texttt{MCBR}, fitted on destriped data. For 70GHz (black), 100GHz (red), 143GHz (green) and 217GHz (blue).}  \label{fig:bias}
\end{center}
\end{figure}

Figure~\ref{fig:bias} shows the bias in $n_{s}$ and $A_{s}$ as a function of the extreme beams fitted on destriped data.
We consider the error on $B_{\ell}$ given by $(r_{\ell} -1)$ for  $\ell=1/\sigma$ representing the sigma of the beam. 
The corresponding values are given in Table~\ref{tab:bias-ns-as}.
For example for 100\,GHz an uncertainty of the extreme beam transfer function $b^{2}_{\ell}$  for  $\ell=810$ of  $\simeq 0.1 \%$  bias the likelihood  by $0.3 \sigma$ and  $0.13 \sigma$ for $n_{s}$ and $A_{s}$ respectively. A beam transfer function known up to $0.02\%$ will bias $n_{s}$ by $0.1 \sigma$. If we had not taken into account the beam uncertainties, then the same deviation in the transfer functions would have biased $n_{s}$ by as much as $0.4 \sigma$, as can be inferred from Figure~\ref{fig:sigmafrac}. 
The inadequacy of a likelihood that does not integrate the beam uncertainties is mentioned in (\cite{kevinetal09}). There a simplified analysis of noisier data (only 1 horn) with all parameters except $n_{s}$ fixed indicated that limiting the bias to $0.1 \sigma$ would require knowledge of $b^{2}_{\ell}$ to 0.04\% where it has fallen to 1\% of peak ($\ell \simeq 1900$ for 100\,GHz).  In our analysis here we see that at $\ell \simeq 1900$ an uncertainty of 0.5\% for the extreme function would bias $n_{s}$ by $0.3 \sigma$, while a mild deviation of the order 0.2\% would produce a bias below $0.05 \sigma$ (see Table~\ref{tab:resultsall}). 
Hence a beam deviation five times that reported in (\cite{kevinetal09}) would bias $n_{s}$ by less than $0.1 \sigma$.  This improvement is mostly due to properly marginalizing over the beam uncertainties via the \texttt{MCBR} method.

\begin{table*}[!htb]
\begin{center}
\begin{tabular}{lcccc}
ch     & $\ell$          &   $r_{\ell}^{2}$         &   $bias/ {\sigma}$ ($n_{s}$)     &  $bias/ {\sigma}$  ($A_{s}$) \\ \hline
70     &579         &1.00217                         & 0.1890                &    0.2846   \\
100   &810        &1.000928                      & 0.3282                   &    0.1240    \\
143   &1141      &1.000982                      & 0.3775                   &  0.1055     \\
217   &1620      &1.00043                        & 0.0929                 &   0.0188 \\
\end{tabular}
\caption{Bias on $n_s$ and $A_{s}$ in units of the error due to the deviation of the extreme function $(r^{ext}_{\ell})^{2}$ at $\ell = 1/\sigma$, after \texttt{MCBR}, fitted on destriped data. For each Planck channel.}
\label{tab:bias-ns-as}
\end{center}
\end{table*}

\section{Conclusions}

We have developed a fast new method, \texttt{MCBR}, to propagate beam uncertainties to parameter estimation.  The method properly accounts for the marginalised errors in the parameters. A desirable feature of the method is that it makes minimal assumptions on beam uncertainties.  For example, it does not assume the data are normally distributed, and, unlike other approaches such as analytic marginalization, it does not require Gaussian priors on the specific systematic uncertainty.  Furthermore it accounts accurately for the shape of the beam as it makes use of beam uncertainty templates for such beams, hence there is no need for simplified a priori assumptions on their shapes.  Finally \texttt{MCBR} can be generalized and used to propagate other systematic uncertainties, as long as a set of templates of such systematics is provided.

From the study presented here on propagating the beam measurement errors to parameter estimation via the new \texttt{MCBR} method for Planck and for a future experiment, we conclude:

\begin{itemize}
\item Removal of $1/f$ noise residuals, by destriping or other techniques, is quite important. 
\item The main impact of  beam uncertainties is to widen the marginal distributions of some parameters (most notably $n_s$). 
\item Assuming as extreme beam deviation in the simulated data results in a biased estimation of some parameters (mainly of $n_s$) due to incomplete marginalization. 
\item The parameters more noticeably impacted by beam uncertainties are:  $n_s$, $\Omega_bh^2$ and $A_s$
\end{itemize}

These results demonstrate the relevance of applying destriping techniques on Planck data to remove 1/f noise.  

When the beam fitting is performed in destriped data the uncertainties on the beams for say 100GHz are at most of the order of $0.5 \%$ for $\ell = 1500$ which translates into an increase of parameter uncertainties at most of the order of $0.1 \%$.  Instead the uncertainties on $A_{s}$  at 70GHz and on $n_s$ at 100GHz increases approximately by $20 \%$ and $26 \%$ respectively for beams fitted on white + $1/f$ noise data while it remains unaltered for white noise background alone. 

The effect of wrong assumptions on beam parameters will bias the parameter constraints only for extreme deviations from the true beam and hence for quite atypical circunstances.
Considering the analysis performed on destriped data,  at 100\,GHz an uncertainty of the extreme beam transfer function at $\ell=810$ of $\simeq 0.1 \%$ will bias the likelihood by $0.3 \sigma$ and $0.13 \sigma$ for $n_{s}$ and $A_{s}$, respectively.  A beam transfer function known to  $0.02\%$ will bias $n_{s}$ by $0.1 \sigma$. If we had not taken into account the beam uncertainties, then the same deviation in the transfer functions would have biased $n_{s}$ by as much as $0.4 \sigma$.
To limit the bias in $n_{s}$ to less than $0.1 \sigma$ will require a knowledge of a mild deviated beam $b_l^2$ to $0.2\%$ where it has fallen to 1 percent. A mild deviated function gives rise to no observable bias (ie at most of the order $0.05 \sigma$).

Therefore we expect only a small impact of beam measurement errors on cosmological parameter estimation as long as the beam fitting is performed on destriped data.

\begin{acknowledgements}
GR is grateful to Jeffrey Jewell and Lloyd Knox for  insightful discussions. LP acknowledges support by ASI contract I/016/07/0 "COFIS". KMH receives support from NASA via JPL subcontract 1363745. 
We gratefully acknowledge support  by the NASA Science Mission Directorate via the US Planck Project. The research described in this paper was partially carried out at the Jet propulsion Laboratory, California Institute of Technology, under a contract with NASA.
Copyright 2009. All rights reserved.
\end{acknowledgements}


\end{document}